\documentclass[useAMS,usenatbib]{mn2e}

\title[MALT-45 - I. Techniques and spectral line data]
  {MALT-45: A 7\,mm survey of the southern Galaxy - I. Techniques and spectral line data}
\author[C. H. Jordan et al.]
  {Christopher~H.~Jordan,$^{1,2}$\thanks{E-mail: christopher.jordan@utas.edu.au}
    Andrew~J.~Walsh,$^3$
    Vicki~Lowe,$^{2,4}$
    Maxim~A.~Voronkov,$^2$\newauthor
    Simon~P.~Ellingsen,$^1$
    Shari~L.~Breen,$^2$
    Cormac~R.~Purcell,$^5$
    Peter~J.~Barnes,$^7$\newauthor
    Michael~G.~Burton,$^4$
    Maria~R.~Cunningham,$^4$
    Tracey~Hill,$^8$
    James~M.~Jackson,$^9$\newauthor
    Steven~N.~Longmore,$^6$
    Nicolas~Peretto$^{10}$
    and James~S.~Urquhart$^{11,2}$\\
  $^1$School of Physical Sciences, Private Bag 37, University of Tasmania, Hobart, Tasmania 7001, Australia\\
  $^2$CSIRO Astronomy and Space Science, PO Box 76, Epping, NSW 1710, Australia\\
  $^3$International Centre for Radio Astronomy Research, Curtin University, GPO Box U1987, Perth, WA 6845, Australia\\
  $^4$School of Physics, University of New South Wales, Sydney, NSW 2052, Australia\\
  $^5$School of Physics, University of Sydney, Sydney, NSW 2006, Australia\\
  $^6$Astrophysics Research Institute, Liverpool John Moores University, IC2, Liverpool Science Park, Liverpool L3 5RF, UK\\
  $^7$Astronomy Department, University of Florida, Gainesville, FL 32611, USA\\
  $^8$Joint ALMA Observatory, Alonso de Cordova 3107, Vitacura, Santiago, Chile\\
  $^9$Astronomy Department, Boston University, 725 Commonwealth Avenue, Boston, MA 02215, USA\\
  $^{10}$School of Physics \& Astronomy, Cardiff University, Queens Buildings, The Parade, Cardiff CF24 3AA, UK\\
  $^{11}$Max-Planck-Institut f\"{u}r Radioastronomie, Auf dem H\"{u}gel 69, Bonn, Germany\\
}

\date{Accepted 2015 January 23. Received 2015 January 22; in original form 2014 December 1}

\pagerange{\pageref{firstpage}--\pageref{lastpage}} \pubyear{2015}
\pagerange{1--19} \pubyear{2015}

\def\LaTeX{L\kern-.36em\raise.3ex\hbox{a}\kern-.15em
    T\kern-.1667em\lower.7ex\hbox{E}\kern-.125emX}

\usepackage{caption}

\usepackage{rotating}
\usepackage{lscape}
\usepackage{longtable}

\usepackage{subfigure}
\usepackage{graphicx}

\usepackage{xspace}

\usepackage{color}
\definecolor{highlight}{rgb}{1,0,0}

\newcommand{\kms}{\mbox{km\,s$^{-1}$}\xspace}

\newcommand{\cs}{CS\xspace}
\newcommand{\choh}{CH$_3$OH\xspace}
\newcommand{\sio}{SiO\xspace}
\newcommand{\nh}{NH$_3$\xspace}
\newcommand{\water}{H$_2$O\xspace}
\newcommand{\co}{CO\xspace}
\newcommand{\oh}{OH\xspace}

\newcommand{\hi}{H\,\textsc{i}\xspace}
\newcommand{\hii}{H\,\textsc{ii}\xspace}
\newcommand{\uchii}{UCH\,\textsc{ii}\xspace}

\newcommand{\chohno}{77\xspace}
\newcommand{\siototal}{47\xspace}

\begin{document}
\label{firstpage}
\maketitle

%%%%%%%%%%%%%%%%%%%%%%%%%%%%%%%%%%%%%%%%%%%%%%%%%%%%%%%%%%%%%%%%%%%%%%%%%%%%%%%%
\begin{abstract}
We present the first results from the MALT-45 (Millimetre Astronomer's Legacy Team - 45\,GHz) Galactic Plane survey. We have observed 5 square-degrees ($l = 330 - 335$, $b = \pm0.5$) for spectral lines in the 7\,mm band (42--44 and 48--49\,GHz), including \cs (1--0), class I \choh masers in the 7(0,7)--6(1,6) A$^{+}$ transition and \sio (1--0) $v=0,1,2,3$. MALT-45 is the first unbiased, large-scale, sensitive spectral line survey in this frequency range. In this paper, we present data from the survey as well as a few intriguing results; rigorous analyses of these science cases are reserved for future publications. Across the survey region, we detected 77 class I \choh masers, of which 58 are new detections, along with many sites of thermal and maser \sio emission and thermal \cs. We found that 35 class I \choh masers were associated with the published locations of class II \choh, \water and \oh masers but 42 have no known masers within 60 arcsec. We compared the MALT-45 \cs with \nh (1,1) to reveal regions of \cs depletion and high opacity, as well as evolved star-forming regions with a high ratio of \cs to \nh. All \sio masers are new detections, and appear to be associated with evolved stars from the {\em Spitzer} Galactic Legacy Infrared Mid-Plane Survey Extraordinaire (GLIMPSE). Generally, within \sio regions of multiple vibrational modes, the intensity decreases as $v=1,2,3$, but there are a few exceptions where $v=2$ is stronger than $v=1$.
\end{abstract}


%%%%%%%%%%%%%%%%%%%%%%%%%%%%%%%%%%%%%%%%%%%%%%%%%%%%%%%%%%%%%%%%%%%%%%%%%%%%%%%%
\section{Introduction}
High-mass stars are critical elements of Galactic structure, due to their evolution in molecular clouds, the turbulent energy they inject into the interstellar medium and the metals dispersed in their death. These stars are categorised by having mass $>8M_{\odot}$ and eventually explode as type II supernovae, transforming the local interstellar medium.

The mechanism of high-mass star formation (HMSF) remains an important unsolved problem in astrophysics. Understanding HMSF is difficult for a number of reasons: the rarity of HMSF regions, high dust extinction within molecular clouds, rapid evolution, and large distances from our Solar System \citep{zinnecker07} all combine to hinder our understanding. HMSF is known to begin within Giant Molecular Clouds (GMCs) and eventually end with star clusters or associations \citep{lada03}. To understand the various stages of HMSF, an inventory of regions must be compiled in a range of known tracers.

To date, there have been many surveys conducted to help understand HMSF. These have been useful in identifying HMSF regions and their characteristic spectral lines, but tend to emphasise subsections of the HMSF timeline. Examples of these surveys include methanol masers (e.g. \citealt{walsh98,cyganowski09,green09}), infrared dark clouds (IRDCs; e.g. \citealt{peretto09}) and radio continuum sources (e.g. \citealt{purcell13}). Methanol masers act as reliable signposts of an early evolutionary stage of HMSF \citep{ellingsen06}, but do not occur at every evolutionary stage; IRDCs are only seen when they are close to us, projected against a bright infrared background and are not always associated with HMSF \citep{kauffmann10}; radio continuum sources are powered by main sequence high-mass stars, and are seen only when their free-free emission has penetrated through the dense surroundings.

Large area surveys of molecular gas are ideal for identifying HMSF regions across a broad range of evolutionary phases. In particular, the critical density of a gas tracer can act as a probe for Galactic structure and star formation. Critical density ($n_c$) is a measure of the density required to produce detectable emission \citep{evans99}. Previous surveys for star formation in external galaxies, typically for low-density gas tracers ($n_c < 10^3$\,cm$^{-3}$), suffer from poor resolution (on the order of kpc) and sensitivity. Untargeted Galactic plane surveys avoid these issues, while simultaneously offering the ability to detect new, less obvious regions of star formation within our Galaxy. Other untargeted surveys have been productive in identifying extended emission, such as in \co \citep{dame01,jackson06,burton13}, \oh \citep{dawson14} and \hi \citep{mcclure-griffiths05}. However, HMSF occurs in regions of dense molecular gas, and so mapping high-density tracers serves well to identify target regions. \co (1--0) has a relatively low effective critical density ($n_c = $ $\sim$10$^2$\,cm$^{-3}$) compared to other tracers, such as \nh (1,1) ($\sim$10$^3$\,cm$^{-3}$; \citealt{evans99}) and HC$_3$N (4--3) ($\sim$10$^4$\,cm$^{-3}$; \citealt{fuller93}), which have been mapped by the \water Galactic Plane survey (HOPS; \citealt{walsh11}). HOPS has been successful in identifying previously unknown sites of star formation, as well as probing the structure of the Milky Way's spiral arms.

Another, even higher density gas tracer useful for detecting HMSF and mapping the structure of our Galaxy is \cs. The ground state transition $J=(1-0)$ for \cs lies within the 7\,mm waveband, and has an effective critical density $n_c = $ $4.6\times10^4$\,cm$^{-3}$ at 10\,K \citep{evans99}. Previous large scale observations towards specific regions, such as RCW 106 \citep{rodgers60} for \cs and other high-critical-density molecules have been productive in HMSF research \citep{lo09}. Untargeted observations of \cs over bright infrared regions such as RCW 106 and regions without infrared emission reveal the full population of star-forming candidates to a small sensitivity limit.

The 7\,mm waveband also contains the 44\,GHz 7(0,7)--6(1,6) A$^{+}$ class I methanol maser line, which is the brightest of the class I species. Previous observations of the class I maser have almost always been targeted (e.g. \citealt{slysh94,voronkov14}), which limits and biases our understanding of this astrophysical phenomenon. The 7\,mm waveband includes other spectral lines detailed in Table \ref{tab:spectral_lines}. To help develop our understanding of HMSF using the diagnostics in the 7\,mm waveband, we have devised the MALT-45 untargeted Galactic plane survey.

\subsection{Spectral lines}
\label{sec:spectral_lines}
\cs has a slightly higher effective critical density compared to \nh ($n_c = $ $4.6\times10^4$\,cm$^{-3}$ vs. $\sim$10$^3$\,cm$^{-3}$), but suffers from freeze-out on to dust grains in the coldest and most dense regions \citep{bergin01}. \nh appears to be resilient to freeze-out in these regions \citep{tafalla02}, therefore by comparing the dense gas tracers from MALT-45 and HOPS, we are able to identify these cold, dense regions in our Galaxy by analysing the \cs/\nh ratio and \nh temperature. MALT-45 simultaneously surveys C$^{34}$S (1--0), which is a less abundant isotopologue of \cs (assumed \cs/C$^{34}$S ratio of 24.4; \citealt{chin96}). A comparison between \cs and C$^{34}$S allows us to measure optical depth at these locations. C$^{34}$S data are not formally presented in this paper, but a preliminary data reduction has been performed for an optical depth investigation in Section \ref{sec:cs_nh3}.

There are two varieties of \choh maser: class I and II. The class II variety is well known and studied; it is radiatively excited by young stellar objects with sufficient mass to be considered HMSF. Thus, it is vitally important as it occurs only in regions of HMSF, but only at a specific evolutionary state \citep{walsh98,minier03,green12,breen13}. In contrast, class I masers are poorly understood. Research has shown that the class I maser is collisionally excited, and associated with shocked gas \citep{voronkov10a,voronkov10b,voronkov14}, but not necessarily only from outflows, as seen in \citet{pihlstrom14}. It is known to occur in both low-mass \citep{kalenskii10} and high-mass regions of star formation \citep{gan13}. Most known class I masers have been discovered in targeted searches towards other sources, such as 6.7\,GHz class II masers \citep{ellingsen05}, HMSF regions with and without \uchii regions \citep{kurtz04} and extended green objects (EGOs; \citealt{cyganowski08,chen11}). Because class I masers have been found towards a range of sources and evolutionary states, their conditions for excitation remain uncertain. By conducting a sensitive and untargeted survey, MALT-45 reveals the full population of class I masers to a small sensitivity limit ($5\sigma$ sensitivity of 4.5\,Jy). This eliminates biases towards known tracers inherent in targeted observations, and helps understand their conditions of excitation.

\sio masers are commonly associated with oxygen rich Asymptotic Giant Branch (AGB) or Red Supergiant (RSG) stars, occurring in the extended atmosphere within a few stellar radii \citep{elitzur92,messineo02}. These masers are useful in determining radial velocities to such stars to within a few \kms \citep{messineo02}. \citet{messineo02} conducted a survey for \sio masers at 86\,GHz towards evolved stars, and demonstrated their use in probing Galactic kinematics through the maser velocities. Additionally, \sio masers have been detected towards star-forming regions, but despite an extensive search towards a wide range of HMSF regions, only three are known (Orion Source I, W51 and Sgr B2; \citealt{zapata09}). \sio maser emission in Orion provides the best evidence of a high-mass accretion disc \citep{matthews07}. MALT-45 has the potential to reveal many new sources of \sio (1--0) maser emission, both towards evolved stars and star-forming regions.

Along with the maser emission, MALT-45 also includes \sio (1--0) $v=0$, which is thermal line emission. Thermal \sio is commonly found in star-forming regions, produced when silicate-bearing grains are destroyed. Thus, it is a good tracer of shocked gas and outflows \citep{martin-pintado92}.

In this paper, we focus on detailing the survey from $330^{\circ} \leq l \leq 335^{\circ}$, $b=\pm0.5^{\circ}$ and present results of the global properties of \cs (1--0), class I \choh masers and \sio (1--0) thermal and maser emission. This paper is one of a series; another paper will be dedicated to maser follow-up observations, and another for clump analysis with \cs data. In this paper, we focus on comparing: \cs emission to \nh and class I \choh masers; class I \choh masers to class II \choh, \water and \oh masers, thermal \sio and infrared emission, and 95\,GHz class I \choh masers; and \sio masers to infrared emission. We emphasise that the science cases presented in this paper are used primarily to highlight a few intriguing results and discuss some qualitatively plausible explanations. Full analyses for these cases will appear in future publications.

%%%%%%%%%%%%%%%%%%%%%%%%%%%%%%%%%%%%%%%%%%%%%%%%%%%%%%%%%%%%%%%%%%%%%%%%%%%%%%%%
\vspace{-3mm}
\section{Survey Design}
MALT-45 is an untargeted Galactic plane survey for spectral lines which are commonly bright in star-forming regions at 45\,GHz (7\,mm waveband). We have so far observed 5 square-degrees within the region bounded by $330^{\circ} \leq l \leq 335^{\circ}$, $b=\pm0.5^{\circ}$. MALT-45 observations were conducted on the Australia Telescope Compact Array (ATCA), which provides 2$\times$ 2048\,MHz broadband continuum windows for observing. Section \ref{sec:spectral_lines} discusses the primary lines surveyed, and their rest frequencies dictate the positions of the broadband windows for MALT-45. Within the frequency ranges of the broadband windows, we survey for 12 spectral lines; see Table \ref{tab:spectral_lines}.

\begin{table*}
  \begin{center}
    \caption{Bright spectral lines mapped by MALT-45 between 42.2 and 49.2\,GHz. The first column lists the spectral line. The second column lists the rest frequency of the line. The third column lists the uncertainty of the rest frequency. The fourth column classifies the line as either a maser or thermal line. The fifth column gives the ATCA pointing beam size at this frequency. The sixth column indicates whether this line is discussed in this paper (``Y'') or not (``N''). Columns seven and eight detail the LSR velocity range covered by the CABB zoom band for this spectral line. The ninth column lists the median RMS noise level per spectral channel, with errors representing the standard deviation. The RMS noise is given to two significant figures. Radio recombination line (RRL) frequencies are taken from \citet{lilley68}. All other rest frequencies are taken from the Cologne Database for Molecular Spectroscopy (CDMS; \citealt{muller05}). The velocity resolution ranges from 0.225\,\kms (lowest rest frequency) to 0.195\,\kms (highest rest frequency) per channel.}
  \label{tab:spectral_lines}
  \vspace{-3mm}
  \begin{tabular}{ lcccc ccccc }
    \hline
    Spectral Line & Frequency & Uncertainty & Maser or & Beam Size & Detailed & \multicolumn{2}{c}{Velocity Range} & Median RMS  \\
                  & (GHz)     & (kHz)       & thermal? & (arcsec)  & in this  & Min & Max                          & Noise Level \\
                  &           &             &          &           & paper?   & \multicolumn{2}{c}{(\kms)}         &             \\
    \hline
    \sio (1--0) $v=3$                & 42.51934 & 1   & Maser   & 66 & Y & -287 & 167 & 0.85$\pm$0.07\,Jy \\
    \sio (1--0) $v=2$                & 42.82048 & 1   & Maser   & 66 & Y & -194 & 257 & 0.82$\pm$0.06\,Jy \\
    H53\,$\alpha$ (RRL) & 42.95197 &     & Thermal & 65 & N & & & & \\
    \sio (1--0) $v=1$                & 43.12203 & 2   & Maser   & 65 & Y & -321 & 127 & 0.83$\pm$0.06\,Jy \\
    \sio (1--0) $v=0$                & 43.42376 & 2   & Thermal & 65 & Y & -224 & 221 & 27$\pm$1.7\,mK \\
    \choh 7(0,7)--6(1,6) A$^{+}$     & 44.06941 & 10  & Maser (Class I)&64&Y&-183&256 & 0.90$\pm$0.09\,Jy \\
    H51\,$\alpha$ (RRL) & 48.15360 &     & Thermal & 58 & N & & & & \\
    C$^{34}$S (1--0)                 & 48.20694 & 2   & Thermal & 58 & N & & & & \\
    \choh 1$_0$--0$_0$ A$^{+}$       & 48.37246 & 0.7 & Thermal & 58 & N & & & & \\
    \choh 1$_0$--0$_0$ E             & 48.37689 & 10  & Thermal & 58 & N & & & & \\
    OCS (4--3)                       & 48.65160 & 1   & Thermal & 58 & N & & & & \\
    \cs (1--0)                       & 48.99095 & 2   & Thermal & 57 & Y & -157 & 237 & 34$\pm$6.8\,mK \\
    \hline
  \end{tabular}
  \end{center}
\end{table*}

\subsection{Australia Telescope Compact Array characterisation}
The ATCA is an interferometer composed of six 22\,m antennas, CA01 through CA06. In recent years, the ATCA correlator has been upgraded to the Compact Array Broadband Backend (CABB; \citealt{wilson11}), which now provides auto-correlation spectra. In addition to this upgrade, the ATCA obtained on-the-fly (OTF) mosaicing capabilities. We are particularly interested in the auto-correlations, as combining the six antennas in this way is similar to observing six times longer with an individual 22\,m antenna such as Mopra. By using the ATCA auto-correlations for 7\,mm surveying, we obtain an excellent metric on sensitivity versus survey speed. See \citet{jordan13} for pilot observations with MALT-45.

For the results presented here, only data from antennas CA01 through CA05 are presented. CA06 has a higher surface root-mean-square (RMS) and poorer sensitivity in the 7\,mm wavelength range. In future observations, it may be possible to include CA06 data by weighting its contribution accordingly, but it is ignored at present.

\subsubsection{CABB zoom modes}
The recent upgrades to the CABB correlator provide the option of observing with ``zoom modes'', which act as high-resolution spectral windows. MALT-45 makes use of the 64M-32k configuration. In addition to the two 2048\,MHz broadband continuum windows, up to 16$\times$64\,MHz ``zoom windows'' can be allocated in each 2048\,MHz range. Each zoom window has 2048 spectral channels, yielding a fine resolution of 32\,kHz. At 45\,GHz, this is approximately 0.2\,\kms per channel. Each spectral line listed in Table \ref{tab:spectral_lines} was observed with a 64M-32k zoom window (with the exception of the thermal \choh 1$_0$--0$_0$ A$^{+}$ and E lines, which are separated by 4\,MHz and were simultaneously observed within one zoom), using a total of 11 zooms. Each zoom window is ideally placed upon the spectral line centre frequency, although the exact placement depends on the way the correlator distributes the zooms. Every zoom provides approximately 400\,\kms of velocity coverage, and is detailed in Table \ref{tab:spectral_lines}.

\subsection{MALT-45 observations}
\begin{figure}
  \includegraphics[width=0.47\textwidth]{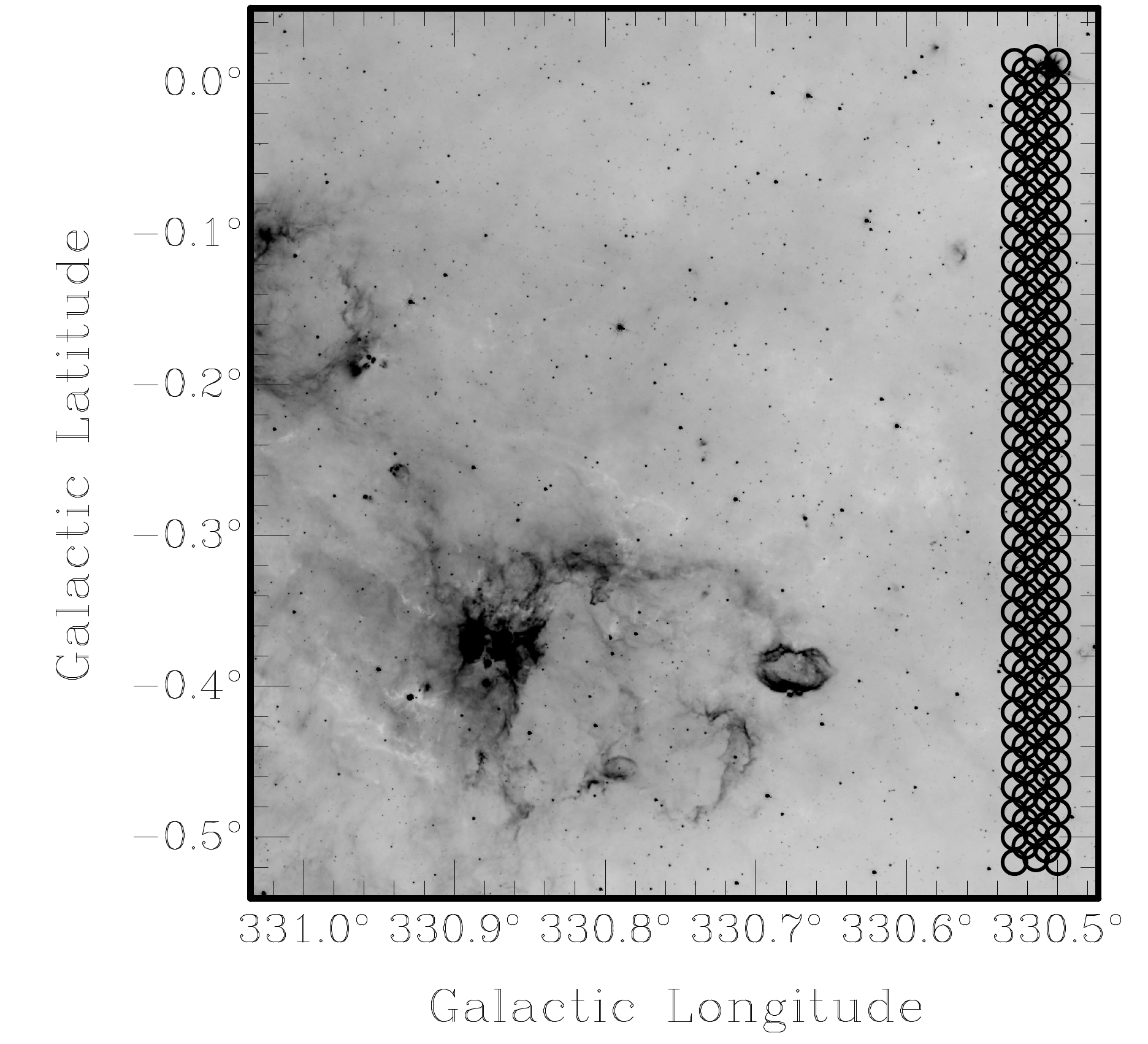}
  \caption{GLIMPSE 8.0\,$\mu$m emission overlaid with part of a MALT-45 mosaic. Each circle represents the primary beam of the ATCA at 49.2\,GHz, and is a pointing of the mosaic. Every quarter-square-degree region mapped by MALT-45 is built up of pointing ``rows'' seen here. These five rows represent the start of the Galactic latitude scans for the G330.5$-$0.5 region. A similar mosaic is made for each region in Galactic longitude. When both Galactic latitude and longitude mosaics are combined, a fully Nyquist sampled map is obtained. Rows are classified as either ``long'' or ``short''; in this figure, the three long rows have two short rows interleaved.}
  \label{fig:mosaic}
\end{figure}

All observations were conducted with the same surveying technique. Scans were observed in quarter-square-degree regions in Galactic coordinates, in both latitude and longitude directions. For both latitude and longitude scans, the mosaics are represented by ``long'' and ``short'' interleaved pointings. See Figure \ref{fig:mosaic} for an illustrative example. Within a mosaic row, consecutive pointing centres are separated by 1.045 beams, and centres of each mosaic row are separated by 0.45 beams. Each beam (pointing) is observed for 6 seconds in OTF mode observing. The frequency used to calculate the size of the primary beam is 49.2\,GHz (the highest observed frequency), thus the beam used was 57.13 arcsec, and separation parameters are 59.7 and 25.7 arcsec, respectively. When combined, the scans provide a fully Nyquist sampled spatial map.

The MALT-45 results presented here were observed over eight months on the ATCA. The first set of observations were conducted on 2012 September 21-23, 28 and 30, and October 1-13. The second set of observations were conducted from 2013 April 10-19. All observations were conducted in the H214 array configuration. The region $330^{\circ} \leq l \leq 333^{\circ}$, $b=\pm0.5^{\circ}$, and part of $333.0^{\circ} \leq l \leq 333.5^{\circ}$, $0^{\circ} \leq b \leq 0.5^{\circ}$ were surveyed in the first observing run. The remainder ($333^{\circ} \leq l \leq 335^{\circ}$, $b=\pm0.5^{\circ}$) was completed during the second observing run. Weather conditions across all observations were generally good, with typical path noises below 300\,$\mu$m. Path noise is a measure of the weather quality, based on the phase delay between successive measurements of a geosynchronous satellite. Values above 500\,$\mu$m are generally considered poor, but conditions only affect interferometric data.

Each quarter square-degree region required approximately 14 hours observation time, including the typical setup and calibration. Pointing corrections were determined using PMN J1646$-$5044 and applied approximately every 70 minutes. The reference position G334.0$-$1.0 was observed every 45 minutes for calibration of the auto-correlation data. In addition, to allow interferometric calibration, bandpass calibration was derived from PKS B1253$-$055, phase calibration from PMN J1646$-$5044 and flux density from PKS 1934$-$63.

\subsection{Data reduction}
Data reduction was undertaken in a similar manner to that of the MALT-45 pilot survey \citep{jordan13}. Auto-correlations are extracted from the raw data using \textsc{MIRIAD}, before being converted into a single dish format. At this stage, the single dish data for each ATCA antenna is treated individually, using standard data reduction tools on a single dish such as Mopra. Each antenna has a baseline removal procedure applied using \textsc{LiveData}, followed by more robust baselining using the ATNF Spectral line Analysis Package (\textsc{ASAP}). \textsc{ASAP} averages each of the antenna auto-correlations together before performing further baseline smoothing. \textsc{ASAP} then provides both XX and YY polarisation products, which are used by \textsc{Gridzilla} to create data cubes.

Flux density calibration was achieved with antenna efficiencies and $T_B$-to-Jy conversion factors taken from \citet{urquhart10}. Using these values assumes that the 7\,mm antenna and receiver response from Mopra is similar to the ATCA. Once the data cubes were gridded, the factors were applied to provide appropriate flux density scaling.

The average RMS noise levels were calculated for each data cube. The median values and their errors are listed in Table \ref{tab:spectral_lines}. Values are calculated by using the \textsc{MIRIAD} task \textsc{imstat}, which determines RMS values for each channel in a cube. Error values represent the standard deviation of all RMS values. RMS noise maps may be seen in Appendix A, Figures 1 through 6. Appendices are available online.

\twocolumn[{
    \begin{minipage}{\textwidth}
      \includegraphics[width=1.0\textwidth]{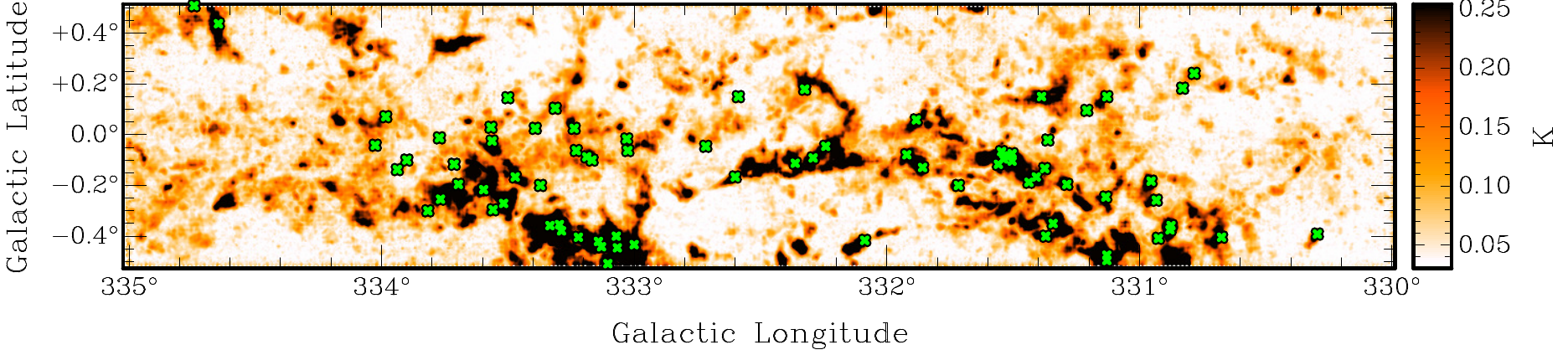}
      \captionof{figure}{Auto-correlated \cs (1--0) peak-intensity map overlaid with positions of detected class I \choh masers (green cross symbols). A main-beam temperature of 0.1\,K represents a $5\sigma$ peak of emission; note that the RMS of this image is less than that stated in Table \ref{tab:spectral_lines} (20\,mK vs. 34mK), because this image has been produced by binning channels into groups of 10.}
      \label{fig:cs_ch3oh}
%--------------------
      \vspace{7mm}
      \includegraphics[width=1.0\textwidth]{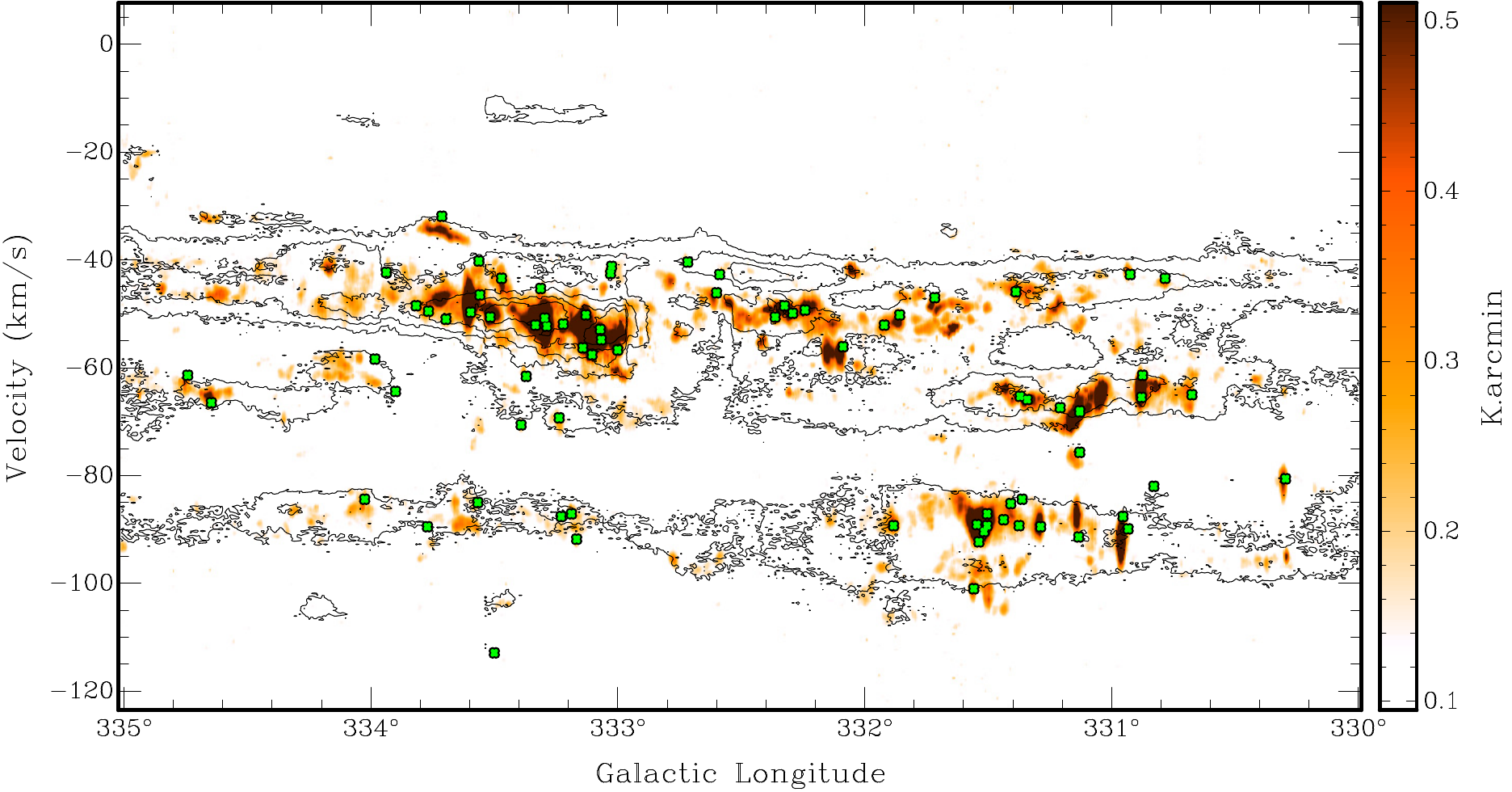}
      \captionof{figure}{\cs (1--0) from MALT-45 overlaid with \co contours from Barnes et al. (ApJ submitted). Contour levels are 90, 180, ...450\,K\,arcmin in units of main-beam temperature, integrated over Galactic latitude. There is excellent agreement in regions of concentrated \co with \cs emission. Green cross symbols represent MALT-45 class I \choh masers, which generally occur in bright regions of \cs; only 4 of the \chohno masers (5 per cent) have a \cs peak intensity less than 0.11\,K ($\sim$3$\sigma$) at their peak position.}
      \label{fig:cs_thrumms}
%--------------------
      \vspace{7mm}
      \includegraphics[width=1.0\textwidth]{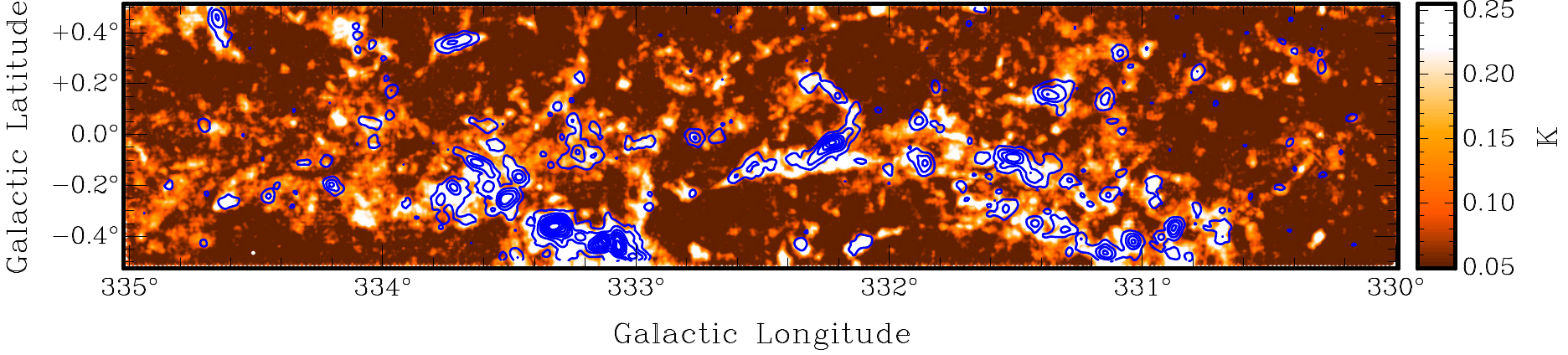}
      \captionof{figure}{Auto-correlated \cs (1--0) peak-intensity map overlaid with HOPS \nh (1,1) peak-signal-to-noise contours. Contour levels are 8, 18, ...88 per cent of 11.13\,K in units of main-beam temperature. The white circle seen at G334.5$-$0.45 is the size of the beam.}
      \label{fig:cs_hops}
      \vspace{10mm}
    \end{minipage}
}]

%%%%%%%%%%%%%%%%%%%%%%%%%%%%%%%%%%%%%%%%%%%%%%%%%%%%%%%%%%%%%%%%%%%%%%%%%%%%%%%%
\section{Results}
For this paper, auto-correlated data cubes were produced for \cs (1--0), class I \choh masers, each of the \sio (1--0) $v=1,2,3$ maser lines and \sio (1--0) $v=0$ thermal emission. Emission was detected in each data cube. The absolute scale of intensity calibration is not well established, and can be affected by systematic effects such as the bandpass subtraction. However, comparing the peak flux densities of MALT-45 class I methanol masers with \citet{voronkov14}, we ascribe 20 per cent as the uncertainty of the intensities in MALT-45 data.

The positions of class I methanol masers from MALT-45 and \citet{voronkov14} were compared to derive a survey astrometric error. The median RMS offset was determined by deriving the offset for each maser site in \citet{voronkov14} and the MALT-45 position. This median is 15 arcsec, and so we ascribe a 15 arcsec astrometric error to MALT-45.

\subsection{\cs (1--0) at 48.990\,GHz}
The most abundant isotopologue of \cs ($^{12}$C$^{32}$S) was detected in extended emission across the entire survey region, as shown in Figures \ref{fig:cs_ch3oh}, \ref{fig:cs_thrumms} and \ref{fig:cs_hops}. Also shown in Figure \ref{fig:cs_hops} is \nh (1,1) emission, as detected in HOPS \citep{purcell12}. Bright \cs emission is largely coincident with HOPS \nh (1,1) emission, but despite having a higher critical density, the \cs appears to be more extended than the \nh; this is discussed further in Section \ref{sec:cs_nh3}. A longitude-velocity-latitude plot reveals a few velocity components to the \cs data, which trace the spiral arms of the Galaxy over the survey region; see Figure \ref{fig:cs_thrumms}.

C$^{34}$S (1--0) at 48.20694\,GHz is also detected across the survey region, but is not detailed in this paper due to an incomplete data reduction. However, a preliminary reduction has been conducted for analysis in Section \ref{sec:cs_nh3}. C$^{34}$S data will be presented in future publications.

\subsection{Class I \choh 7(0,7)--6(1,6) A$^{+}$ masers at 44.069\,GHz}
\label{sec:results_ch3oh}
Within the survey region, \chohno class I \choh masers were detected. Class I \choh maser catalogues from \citet{slysh94,valtts00,ellingsen05} and \citet{chen11} were used to determine previously known masers. All 19 previously known masers were detected by MALT-45. Note that these articles report findings of 95\,GHz class I \choh masers, rather than 44\,GHz masers (except \citealt{slysh94}); if a 44\,GHz maser has not been reported in a position where a 95\,GHz maser has been detected, we still deem this class I maser as a previously known detection. Therefore, 58 of the 77 detections are previously unknown (hereinafter new masers). A histogram of these populations can be seen in Figure \ref{fig:ch3oh_hist}, revealing the number of new masers for various peak flux densities. Images showing the locations of class I \choh masers with respect to thermal molecular gas are shown in Figures \ref{fig:cs_ch3oh}, \ref{fig:cs_thrumms} and \ref{fig:ch3oh_glimpse}. The properties of each class I maser spectrum is classified in Table \ref{tab:ch3oh}. Each un-smoothed maser spectrum can be seen in Appendix B, and a peak intensity map is presented in Appendix C. Appendices are available online.

Masers were identified by visual inspection of the data by analysing various binned versions of the cube, such as 0.4, 0.6, 0.8, 1.0 and 2.0\,\kms bins. This is effective at highlighting weak but broad maser emission; see for example G330.83$+$0.18 in Appendix B (available online). Any peak of emission exceeding $3\sigma$ in the various binned cubes with at least two adjacent channels of $2\sigma$ was considered for candidacy. The position of each maser reflects the brightest pixel within the neighbourhood of bright maser pixels. We compared the velocity of weaker maser candidates with \cs and \sio $v=0$ emission at the same location to strengthen their candidacy. In future work, we will re-observe each maser candidate to greater sensitivity and spatial resolution to confirm (or otherwise) its existence. The velocity range of each maser is determined by a $1.5\sigma$ cutoff in intensity from the peak velocity. As the channel resolution is approximately 0.2\,\kms, velocities are specified to 0.1\,\kms. All flux densities are specified with two significant figures. Intrinsic variation of masers along with calibration uncertainties place peak flux density errors within 20 per cent.

Using the program supplied by \citet{reid09}, kinematic distances have been calculated and are included for each maser in Table \ref{tab:ch3oh}. Parameters used for determining revised velocities and kinematic distances were taken from \citet{green11}, and the $V_{LSR}$ uncertainty from \citet{voronkov14}: $\Theta_0=246$\,\kms, $R_\odot=8.4$\,kpc, $U_\odot=11.1$\,\kms, $V_\odot=12.2$\,\kms, $W_\odot=7.25$\,\kms, $U_s=0$\,\kms, $V_s=-15.0$\,\kms, $W_s=0$\,\kms, $\sigma(V_{LSR})=3.32$\,\kms. All distances are assumed to be kinematically ``near'', unless the distance prescribed by \citet{green11} is ``far''.

\noindent\begin{minipage}{0.47\textwidth}
  \includegraphics[width=\textwidth]{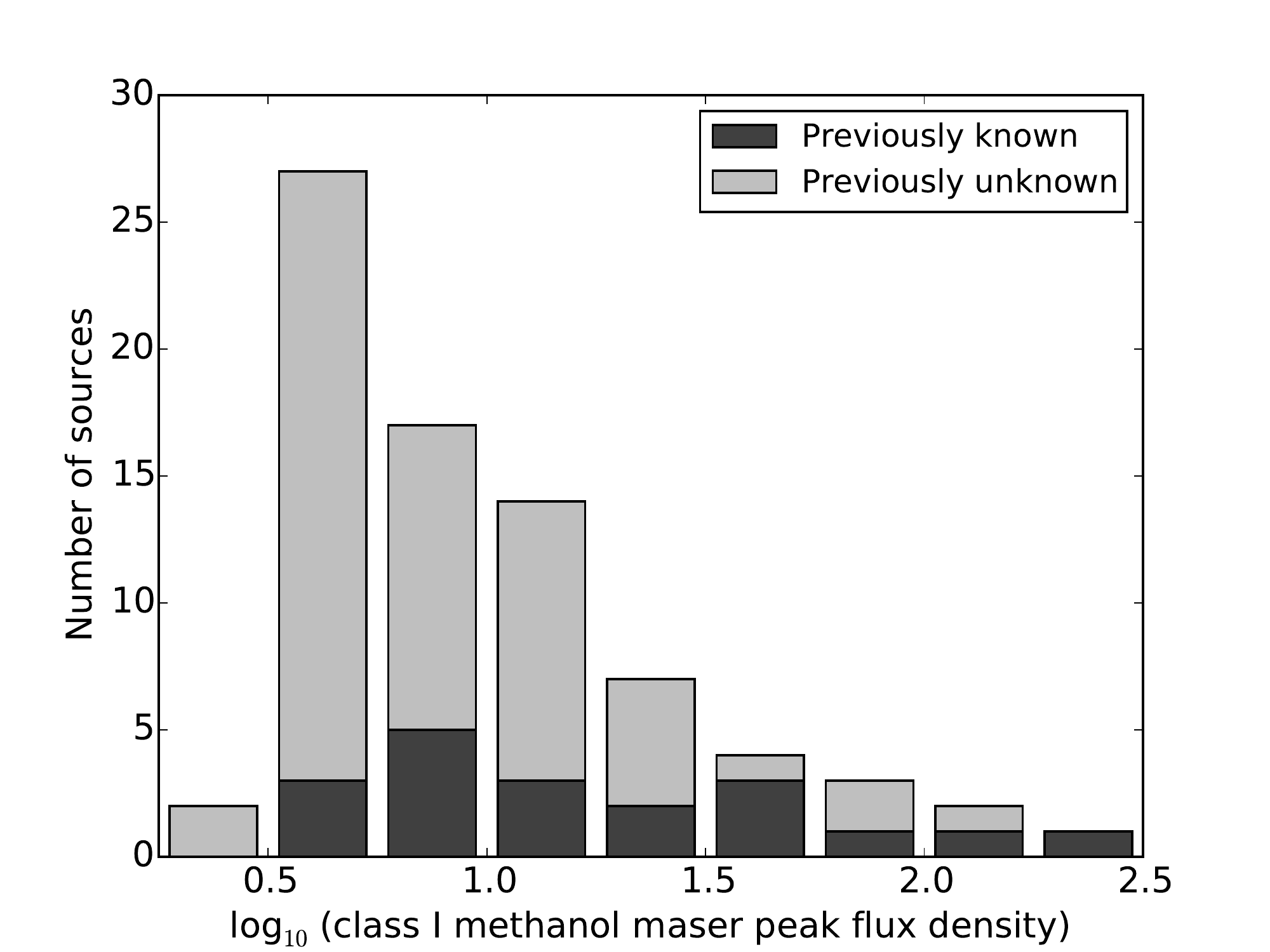}
  \captionof{figure}{A stacked histogram of known and new class I \choh maser peak flux densities. Previously known masers have a near even spread of flux densities, but the new MALT-45 population favours the weaker bins.}
  \label{fig:ch3oh_hist}
\end{minipage}

%% methanol maser detail table
\begin{table*}
  \vspace{-5mm}
  \caption{Properties of detected MALT-45 class I \choh maser emission. Each maser was detected within a 64\,MHz zoom window. The first column lists the source name, derived from Galactic coordinates of the brightest peak. The second and third columns give the coordinates of the maser. The fourth column density lists the source peak flux density. The fifth column lists the RMS noise level of the peak. The sixth and seventh columns list the velocity range of maser emission, within a $1.5\sigma$ cutoff. The eight column lists the velocity at the peak. The ninth column lists kinematic distance calculated, based on position and peak velocity. The tenth and eleventh columns list the errors for the kinematic distance. The twelfth and thirteenth columns list the integrated intensity of \cs (1--0) and \sio (1--0) $v=0$ at the brightest peak. The fourteenth column gives additional remarks, mentioned in the footnotes$^1$. The fifteenth column lists associations with other masers and EGOs, and is detailed in the footnotes$^2$. Association is credited for any presence within 60 arcsec of the class I \choh maser peak position. All velocities are specified to 0.1\,\kms, and all flux densities are specified to two significant figures. Kinematic distances are derived using the kinematic distance model of \citet{reid09}. All distances are assumed to be kinematically ``near'', unless labelled otherwise.}
  \tiny
  \begin{tabular}{ lcccc ccccc ccccc }
      \hline
Source       & RA         & Dec.                                   & Peak    & RMS   & \multicolumn{3}{c}{Velocity} & Kinematic & \multicolumn{2}{c}{Distance} & \multicolumn{2}{c}{Integrated} & Remarks$^1$ & Other     \\
name         & (J2000)    & (J2000)                                & flux    & noise & Min. & Max. & Peak           & distance  & \multicolumn{2}{c}{error}    & \multicolumn{2}{c}{intensity}  &             & masers    \\
             & (h m s)    & ($^\circ$ $^\prime$ $^{\prime\prime}$) & density & level & \multicolumn{3}{c}{(\kms)}   & (kpc)     & \multicolumn{2}{c}{(kpc)}    & \multicolumn{2}{c}{(K\,\kms)}  &             & and       \\
             &            &                                        & (Jy)    &       & & &                          &           & &                            & \cs & \sio                     &             & EGOs$^2$  \\
    \hline
G330.30$-$0.39 & 16 07 37.8 & -52 30 52.1 & 4.4 & 4.9 & -80.8 & -79.5 & -80.6 & 4.90 & 0.16 & -0.16 & 3.1 &  &  & W \\
G330.67$-$0.40 & 16 09 31.2 & -52 16 03.9 & 5.3 & 5.6 & -65.9 & -64.6 & -65.0 & 4.16 & 0.15 & -0.15 & 2.5 &  &  &  \\
G330.78$+$0.24 & 16 07 12.8 & -51 43 03.4 & 3.7 & 4.3 & -43.8 & -43.1 & -43.4 & 3.12 & 0.17 & -0.17 & 1.2 &  &  &  \\
G330.83$+$0.18 & 16 07 41.1 & -51 43 48.0 & 3.1 & 3.8 & -82.3 & -79.7 & -82.0 & 4.95 & 0.16 & -0.16 &  &  &  &  \\
G330.87$-$0.36 & 16 10 16.6 & -52 05 54.3 & 10 & 15 & -67.0 & -58.0 & -61.4 & 4.00 & 0.15 & -0.16 & 8.0 & 0.42 & C &  \\
G330.88$-$0.38 & 16 10 22.3 & -52 06 28.1 & 7.8 & 10 & -67.4 & -58.0 & -65.5 & 4.19 & 0.15 & -0.15 & 8.6 & 0.56 & C & MWSCG \\
G330.92$-$0.41 & 16 10 44.3 & -52 06 04.0 & 28 & 33 & -44.0 & -41.0 & -42.7 & 3.09 & 0.17 & -0.17 & 1.4 &  &  &  \\
G330.93$-$0.26 & 16 10 06.9 & -51 59 07.3 & 5.0 & 5.9 & -91.4 & -89.3 & -89.9 & 5.33 & 0.17 & -0.16 & 1.4 &  &  &  \\
G330.95$-$0.18 & 16 09 53.1 & -51 54 57.9 & 4.0 & 4.7 & -94.8 & -86.7 & -87.6 & 5.21 & 0.16 & -0.16 & 9.6 & 1.4 &  & MWCG \\
G331.13$-$0.48 & 16 12 00.3 & -52 00 38.5 & 5.6 & 7.5 & -69.1 & -66.7 & -68.0 & 4.30 & 0.15 & -0.15 & 5.3 & 0.15 &  &  \\
G331.13$-$0.50 & 16 12 05.9 & -52 01 33.3 & 8.6 & 11 & -68.7 & -67.8 & -68.2 & 4.31 & 0.15 & -0.15 & 4.5 &  &  &  \\
G331.13$-$0.25 & 16 11 00.7 & -51 50 24.8 & 71 & 88 & -92.9 & -82.5 & -91.4 & 5.39 & 0.17 & -0.16 & 4.4 & 0.49 & SVE & MWCG \\
G331.13$+$0.15 & 16 09 15.5 & -51 33 08.1 & 68 & 60 & -79.1 & -74.6 & -75.7 & 4.65 & 0.15 & -0.15 & 1.2 &  &  & MW \\
G331.21$+$0.10 & 16 09 52.2 & -51 32 18.6 & 3.5 & 3.3 & -67.8 & -67.0 & -67.4 & 4.27 & 0.15 & -0.15 &  &  &  &  \\
G331.29$-$0.20 & 16 11 31.3 & -51 41 54.8 & 15 & 20 & -95.7 & -85.0 & -89.5 & 5.29 & 0.16 & -0.16 & 3.7 & 0.17 &  & MWC \\
G331.34$-$0.35 & 16 12 27.4 & -51 46 26.9 & 32 & 37 & -67.2 & -65.0 & -65.9 & 4.21 & 0.15 & -0.15 & 2.1 &  & SVE & MWCG \\
G331.36$-$0.02 & 16 11 06.4 & -51 31 09.7 & 4.7 & 5.8 & -84.6 & -84.2 & -84.4 & 5.05 & 0.16 & -0.15 &  &  &  &  \\
G331.37$-$0.40 & 16 12 48.9 & -51 47 26.2 & 5.9 & 7.3 & -65.9 & -65.0 & -65.3 & 4.18 & 0.15 & -0.15 & 1.6 &  & C & G \\
G331.37$-$0.13 & 16 11 39.5 & -51 35 35.3 & 3.7 & 4.5 & -89.5 & -88.8 & -89.3 & 5.28 & 0.16 & -0.16 & 1.6 &  &  &  \\
G331.39$+$0.15 & 16 10 28.7 & -51 22 37.6 & 6.5 & 5.8 & -46.5 & -45.5 & -45.9 & 3.26 & 0.16 & -0.17 & 2.5 & 0.19 &  &  \\
G331.41$-$0.17 & 16 11 57.8 & -51 35 40.6 & 4.9 & 6.6 & -86.3 & -84.8 & -85.2 & 5.08 & 0.16 & -0.15 & 2.7 &  &  &  \\
G331.44$-$0.19 & 16 12 11.4 & -51 35 24.1 & 14 & 18 & -92.5 & -86.3 & -88.2 & 9.53 & 0.16 & -0.16 & 2.3 & 0.25 & VF & MWC \\
G331.50$-$0.08 & 16 12 00.6 & -51 27 44.3 & 38 & 49 & -92.7 & -86.7 & -89.3 & 5.27 & 0.16 & -0.16 & 4.1 & 0.14 & P3 &  \\
G331.50$-$0.10 & 16 12 08.2 & -51 29 01.0 & 11 & 14 & -101.8 & -86.1 & -87.1 & 5.17 & 0.16 & -0.15 & 6.5 & 0.31 &  & WSC \\
G331.52$-$0.08 & 16 12 06.3 & -51 27 35.5 & 20 & 18 & -94.0 & -88.0 & -90.4 & 5.33 & 0.16 & -0.16 & 5.8 & 0.13 & P1 &  \\
G331.54$-$0.10 & 16 12 15.4 & -51 27 17.1 & 77 & 77 & -94.2 & -89.1 & -92.3 & 5.42 & 0.17 & -0.16 & 7.4 & 0.29 & P2 &  \\
G331.55$-$0.07 & 16 12 10.1 & -51 25 39.9 & 4.9 & 5.3 & -92.9 & -86.3 & -89.1 & 5.26 & 0.16 & -0.16 & 4.9 & 0.15 &  & MC \\
G331.56$-$0.12 & 16 12 26.8 & -51 27 20.4 & 14 & 16 & -104.8 & -98.2 & -101.0 & 5.87 & 0.19 & -0.18 & 2.9 &  &  & MWC \\
G331.72$-$0.20 & 16 13 33.1 & -51 24 28.0 & 4.1 & 3.8 & -47.2 & -46.8 & -47.0 & 3.32 & 0.16 & -0.17 & 2.0 &  &  &  \\
G331.86$-$0.13 & 16 13 54.0 & -51 15 32.1 & 5.6 & 6.5 & -51.4 & -49.5 & -50.2 & 3.48 & 0.16 & -0.16 & 2.6 & 0.44 &  &  \\
G331.88$+$0.06 & 16 13 11.6 & -51 06 19.8 & 10 & 13 & -90.1 & -85.9 & -89.3 & 5.26 & 0.16 & -0.15 & 2.4 & 0.45 &  &  \\
G331.92$-$0.08 & 16 13 58.2 & -51 10 46.7 & 5.8 & 6.4 & -52.7 & -51.6 & -52.1 & 3.57 & 0.16 & -0.16 & 1.5 &  &  &  \\
G332.09$-$0.42 & 16 16 13.7 & -51 18 30.7 & 17 & 21 & -59.5 & -55.5 & -56.1 & 3.76 & 0.15 & -0.16 & 3.6 &  &  & MW \\
G332.24$-$0.05 & 16 15 18.1 & -50 56 03.2 & 140 & 180 & -51.2 & -45.9 & -49.3 & 3.44 & 0.16 & -0.16 & 3.4 & 0.32 &  &  \\
G332.30$-$0.09 & 16 15 43.9 & -50 55 57.8 & 41 & 37 & -52.5 & -45.9 & -49.9 & 3.47 & 0.16 & -0.16 & 3.0 &  & VE & MWG \\
G332.32$+$0.18 & 16 14 42.0 & -50 42 50.2 & 9.3 & 11 & -51.0 & -48.0 & -48.5 & 3.40 & 0.16 & -0.16 & 1.2 &  &  & W \\
G332.36$-$0.11 & 16 16 08.8 & -50 53 55.2 & 9.8 & 12 & -51.2 & -49.9 & -50.6 & 3.51 & 0.16 & -0.16 & 1.7 &  & C & MWCG \\
G332.59$+$0.15 & 16 16 01.3 & -50 33 12.6 & 3.4 & 4.0 & -43.6 & -42.5 & -42.7 & 11.79 & 0.17 & -0.17 &  &  & CF & MG \\
G332.60$-$0.17 & 16 17 28.1 & -50 46 21.2 & 15 & 18 & -47.6 & -45.5 & -46.1 & 3.29 & 0.16 & -0.17 & 2.2 &  & V & MWG \\
G332.72$-$0.05 & 16 17 27.9 & -50 36 16.5 & 6.4 & 7.3 & -41.0 & -39.1 & -40.4 & 3.01 & 0.17 & -0.17 &  &  &  &  \\
G333.00$-$0.43 & 16 20 27.1 & -50 40 58.3 & 6.6 & 9.8 & -57.2 & -56.1 & -56.7 & 3.80 & 0.15 & -0.15 & 4.8 &  &  & W \\
G333.02$-$0.06 & 16 18 55.5 & -50 24 04.3 & 6.8 & 7.9 & -41.7 & -39.7 & -41.2 & 3.06 & 0.17 & -0.17 & 1.2 &  & E & MW \\
G333.03$-$0.02 & 16 18 44.6 & -50 21 56.0 & 5.6 & 7.0 & -42.9 & -42.1 & -42.5 & 3.12 & 0.17 & -0.17 &  &  &  & M \\
G333.07$-$0.44 & 16 20 48.4 & -50 38 41.3 & 2.7 & 3.2 & -55.3 & -54.8 & -54.8 & 3.72 & 0.15 & -0.16 & 6.5 & 0.20 &  & MW \\
G333.07$-$0.40 & 16 20 37.3 & -50 36 33.5 & 4.1 & 4.6 & -54.4 & -52.7 & -52.9 & 3.63 & 0.15 & -0.16 & 4.2 & 0.21 &  &  \\
G333.10$-$0.51 & 16 21 15.1 & -50 39 45.6 & 14 & 8.1 & -58.9 & -57.0 & -57.6 & 3.85 & 0.15 & -0.15 & 4.2 &  &  & M \\
G333.13$-$0.44 & 16 21 04.0 & -50 35 52.0 & 120 & 150 & -57.2 & -45.5 & -50.2 & 3.50 & 0.16 & -0.16 & 11 & 0.64 & SVE P2 & MWC \\
G333.14$-$0.42 & 16 21 01.8 & -50 34 27.1 & 16 & 18 & -56.7 & -45.9 & -56.3 & 3.79 & 0.15 & -0.15 & 10 & 0.51 & P1 & W \\
G333.16$-$0.10 & 16 19 43.4 & -50 19 43.0 & 5.6 & 7.7 & -92.9 & -91.2 & -91.8 & 5.33 & 0.15 & -0.14 & 1.3 &  & E & M \\
G333.18$-$0.09 & 16 19 45.7 & -50 18 18.3 & 5.6 & 7.1 & -88.4 & -85.7 & -87.1 & 5.13 & 0.14 & -0.14 & 1.7 &  & E & MG \\
G333.22$-$0.40 & 16 21 18.6 & -50 30 23.5 & 17 & 19 & -55.3 & -51.0 & -51.9 & 3.59 & 0.15 & -0.16 & 2.7 &  &  &  \\
G333.23$-$0.06 & 16 19 50.3 & -50 15 28.9 & 250 & 310 & -92.3 & -84.4 & -87.6 & 5.15 & 0.14 & -0.14 & 1.0 & 0.25 & SVE & MWSC \\
G333.24$+$0.02 & 16 19 29.5 & -50 11 23.5 & 3.9 & 5.4 & -69.5 & -67.8 & -69.3 & 4.37 & 0.14 & -0.14 & 1.4 &  &  &  \\
G333.29$-$0.38 & 16 21 29.7 & -50 26 30.3 & 5.4 & 6.3 & -53.6 & -50.6 & -52.3 & 3.61 & 0.15 & -0.16 & 11 & 0.27 &  &  \\
G333.30$-$0.35 & 16 21 25.3 & -50 25 05.4 & 3.5 & 3.6 & -53.1 & -49.5 & -50.8 & 3.54 & 0.16 & -0.16 & 9.8 & 0.18 &  &  \\
G333.31$+$0.10 & 16 19 28.7 & -50 04 50.9 & 7.7 & 13 & -50.6 & -43.8 & -45.3 & 11.74 & 0.17 & -0.16 & 2.1 & 0.12 & ECF & MWG \\
G333.33$-$0.36 & 16 21 36.4 & -50 23 40.6 & 22 & 22 & -53.8 & -49.5 & -52.1 & 3.60 & 0.15 & -0.16 & 6.0 & 0.27 &  &  \\
G333.37$-$0.20 & 16 21 04.4 & -50 15 21.6 & 13 & 16 & -62.3 & -57.6 & -61.6 & 4.03 & 0.15 & -0.15 &  &  &  & W \\
G333.39$+$0.02 & 16 20 10.6 & -50 04 53.7 & 4.1 & 6.5 & -72.9 & -69.7 & -70.6 & 10.60 & 0.14 & -0.14 &  &  & F & MWC \\
G333.47$-$0.16 & 16 21 22.1 & -50 09 42.5 & 21 & 23 & -46.1 & -41.4 & -43.4 & 3.18 & 0.16 & -0.17 & 3.6 &  & E & MWCG \\
G333.50$+$0.15 & 16 20 07.6 & -49 55 10.1 & 3.5 & 5.9 & -114.4 & -112.0 & -112.9 & 6.35 & 0.21 & -0.19 &  &  &  &  \\
G333.52$-$0.27 & 16 22 01.9 & -50 12 11.2 & 6.4 & 8.7 & -51.2 & -50.2 & -50.6 & 3.53 & 0.16 & -0.16 & 3.5 &  &  &  \\
G333.56$-$0.30 & 16 22 19.6 & -50 11 28.6 & 7.8 & 11 & -48.0 & -45.9 & -46.5 & 3.33 & 0.16 & -0.17 & 2.4 &  &  &  \\
G333.56$-$0.02 & 16 21 09.0 & -49 59 48.3 & 54 & 63 & -41.7 & -39.1 & -40.2 & 12.02 & 0.17 & -0.17 & 1.8 &  & EF & M \\
G333.57$+$0.03 & 16 20 55.9 & -49 57 19.4 & 6.2 & 6.4 & -85.7 & -84.8 & -85.0 & 5.04 & 0.14 & -0.14 & 0.95 &  &  &  \\
G333.59$-$0.21 & 16 22 08.5 & -50 06 31.7 & 26 & 34 & -51.2 & -47.6 & -49.7 & 3.49 & 0.16 & -0.16 & 11 & 0.31 & SV & WSC \\
G333.70$-$0.20 & 16 22 29.4 & -50 01 23.8 & 4.4 & 6.1 & -51.2 & -50.8 & -51.0 & 3.55 & 0.16 & -0.16 & 2.4 &  &  &  \\
G333.71$-$0.12 & 16 22 12.8 & -49 57 20.1 & 6.6 & 6.9 & -32.3 & -31.2 & -31.9 & 2.58 & 0.18 & -0.19 &  &  &  &  \\
G333.77$-$0.01 & 16 22 00.7 & -49 50 26.5 & 16 & 20 & -90.5 & -88.8 & -89.5 & 5.22 & 0.14 & -0.14 & 2.3 &  &  &  \\
G333.77$-$0.25 & 16 23 03.5 & -50 00 51.0 & 6.4 & 8.0 & -50.2 & -49.3 & -49.5 & 3.49 & 0.16 & -0.16 & 2.4 & 0.12 &  &  \\
G333.82$-$0.30 & 16 23 28.8 & -50 00 39.2 & 19 & 24 & -49.5 & -47.8 & -48.5 & 3.44 & 0.16 & -0.16 & 1.3 &  &  &  \\
G333.90$-$0.10 & 16 22 57.7 & -49 48 39.2 & 3.4 & 4.1 & -65.3 & -62.9 & -64.4 & 10.92 & 0.15 & -0.14 &  &  & F & M \\
G333.94$-$0.14 & 16 23 17.4 & -49 48 38.4 & 4.6 & 4.9 & -43.1 & -42.3 & -42.3 & 3.14 & 0.17 & -0.17 &  &  &  & MW \\
G333.98$+$0.07 & 16 22 34.6 & -49 37 52.8 & 4.5 & 5.7 & -60.8 & -58.0 & -58.4 & 3.90 & 0.15 & -0.15 &  &  &  &  \\
G334.03$-$0.04 & 16 23 15.0 & -49 40 51.7 & 11 & 14 & -84.8 & -84.0 & -84.4 & 5.01 & 0.14 & -0.14 & 2.0 &  &  &  \\
G334.64$+$0.44 & 16 23 50.4 & -48 54 08.7 & 5.7 & 6.4 & -67.2 & -66.5 & -66.5 & 4.27 & 0.14 & -0.14 & 2.2 &  &  &  \\
G334.74$+$0.51 & 16 23 56.6 & -48 47 03.7 & 18 & 6.7 & -65.5 & -61.4 & -61.4 & 4.05 & 0.14 & -0.15 & 1.1 & 0.17 &  &  \\
    \hline
  \end{tabular}
  \label{tab:ch3oh}
  \flushleft\small
  \vspace{-3mm}
$^1$S - 44\,GHz detection from \citet{slysh94}; V - 95\,GHz detection from \citet{valtts00}; E - 95\,GHz detection from \citet{ellingsen05}; C - 95\,GHz detection from \citet{chen11}; P - indicates a blended spectrum of emission. P1 contains a strong peak of emission at a spatially distinct position, with a different velocity to nearby strong masers (but not as strong as the neighbour), P2 is the strong neighbour, and P3 is the next strongest neighbour; F - this maser has been labelled as being kinematically ``far'', due to a class II \choh maser being labeled as such by \citet{green11}. Note that a source without a S, V, E or C remark is a new detection.
$^2$M - presence of a 6.7\,GHz class II methanol maser \citep{caswell11}; W - presence of a 22\,GHz \water maser \citep{walsh11,walsh14,breen10}; G - presence of an EGO classified by \citet{cyganowski08}; S - presence of a 1612 \oh maser \citep{sevenster97}; C - presence of a 1665 or 1667\,MHz \oh maser \citep{caswell98}.
\end{table*}

\onecolumn
\begin{landscape}
\begin{table*}
  \tiny
  \caption{Properties of detected MALT-45 \sio (1--0) maser emission. Each maser was detected within a 64\,MHz zoom window. The first column lists the source name, derived from Galactic coordinates of the brightest peak. The second and third columns give the coordinates of the maser. Columns four through fifteen are categorised into $v=1$, $v=2$ and $v=3$. For each category, there are four columns, listing the peak flux density within this vibrational mode (if detected), and the minimum, maximum and peak velocities. The sixteenth column lists the RMS noise level, which indicates the most statistically significant maser peak detected among the three modes. The seventeenth column lists the kinematic distance calculated, based on position and peak velocity. The eighteenth and nineteenth columns list the errors for the kinematic distance. The twentieth column gives additional remarks, clarified in the footnotes$^1$. If data are not available in one of the lines for a source, it was not detected by MALT-45. Velocity ranges (between minimum and maximum) are determined by a $1.5\sigma$ cutoff. All velocities are specified to 0.1\,\kms, and all flux densities are specified to two significant figures. Association is credited if the source position is within 30 arcsec of the peak \sio maser position. Kinematic distances are derived using the kinematic distance model of \citet{reid09}. All distances are assumed to be kinematically ``near'', except for G330.14$-$0.39, G331.08$+$0.17, G332.40$-$0.33 and G334.20$+$0.22; when assumed to be near, unphysical distances and errors are generated.}
  \begin{tabular}{ lcccc ccccc ccccc ccccc }
    \hline
Source & RA      & Dec.                                   & \multicolumn{4}{c}{\sio $v=1$ properties} & \multicolumn{4}{c}{\sio $v=2$ properties}    & \multicolumn{4}{c}{\sio $v=3$ properties} & RMS & Kinematic & \multicolumn{2}{c}{Distance} & Remarks$^1$ \\
name   & (J2000) & (J2000)                                & Peak    & \multicolumn{3}{c}{Velocity}    & Peak    & \multicolumn{3}{c}{Velocity}       & Peak    & \multicolumn{3}{c}{Velocity}    & noise  & distance  & \multicolumn{2}{c}{error}    & \\
       & (h m s) & ($^\circ$ $^\prime$ $^{\prime\prime}$) & flux    & Min. & Max. & Peak              & flux    & Min. & Max. & Peak & flux & Min. & Max. & Peak              & level & (kpc)     & \multicolumn{2}{c}{(kpc)}    & \\
       &         &                                        & density & \multicolumn{3}{c}{(\kms)}      & density & \multicolumn{3}{c}{(\kms)} & density & \multicolumn{3}{c}{(\kms)}      &       &           & & & \\
       &         &                                        & (Jy)    & & & & (Jy) & & & & (Jy)    & & & & & & & & \\
    \hline
G330.12$-$0.31 & 16 06 24.3 & -52 34 33.5 &  &  &  &  & 3.5 & -65.8 & -61.2 & -61.4 &  &  &  &  & 4.8 &  3.46 &  0.23 & -0.23 & \\
G330.14$-$0.39 & 16 06 52.6 & -52 37 04.9 & 6.3 & -2.6 & 4.5 & 3.5 & 7.5 & -3.0 & 5.5 & 3.6 &  &  &  &  & 11 &  5.05 &  0.45 & -0.41 & W \\
G330.47$+$0.03 & 16 06 36.7 & -52 05 15.6 & 23 & -59.5 & -48.3 & -53.7 & 13 & -59.7 & -51.6 & -54.4 & 2.5 & -56 & -52.7 & -53.8 & 26 &  3.12 &  0.24 & -0.24 & W \\
G330.51$+$0.01 & 16 06 53.2 & -52 04 19.6 & 11 & -13.7 & -9.8 & -12.2 & 19 & -15.5 & -10.2 & -12.8 &  &  &  &  & 19 &  0.76 &  0.34 & -0.36 & \\
G330.55$+$0.15 & 16 06 28.1 & -51 56 41.6 & 4.0 & -68.7 & -67.8 & -68.0 & 2.7 & -69.7 & -68 & -69.3 &  &  &  &  & 4.9 &  3.80 &  0.22 & -0.23 & \\
G330.75$+$0.20 & 16 07 15.2 & -51 46 25.6 & 2.9 & -66.1 & -65.2 & -65.8 & 2.7 & -64.0 & -67.1 & -63.8 &  &  &  &  & 3.8 &  3.63 &  0.22 & -0.23 & \\
G330.85$-$0.44 & 16 10 29.3 & -52 10 34.9 & 3.4 & -32.0 & -30.7 & -31.7 &  &  &  &  &  &  &  &  & 4.3 &  1.98 &  0.28 & -0.29 & W \\
G331.08$+$0.17 & 16 08 55.8 & -51 34 14.6 & 4.9 & -6.1 & -3.9 & -5.2 & 3.8 & -6.9 & -3.7 & -5.8 &  &  &  &  & 7.3 &  4.45 &  0.40 & -0.37 & \\
G331.13$-$0.27 & 16 11 06.2 & -51 51 19.7 & 4.6 & -31.5 & -30.0 & -30.7 & 4.7 & -32.1 & -30.3 & -31.0 &  &  &  &  & 7.7 &  1.94 &  0.28 & -0.29 & \\
G331.21$+$0.25 & 16 09 10.8 & -51 25 19.1 & 2.9 & -36.7 & -34.6 & -36.1 & 3.0 & -37.1 & -36.0 & -36.3 &  &  &  &  & 4.4 &  2.24 &  0.27 & -0.28 & \\
G331.24$-$0.02 & 16 10 32.2 & -51 36 05.6 & 2.7 & -90.0 & -89.1 & -89.1 &  &  &  &  &  &  &  &  & 3.2 &  4.69 &  0.22 & -0.21 & \\
G331.48$-$0.03 & 16 11 43.7 & -51 26 46.0 & 7.0 & -69.5 & -66.3 & -68.0 & 5.7 & -69.7 & -62.9 & -68.4 &  &  &  &  & 9.1 &  3.80 &  0.22 & -0.22 & \\
G331.58$+$0.07 & 16 11 44.5 & -51 18 06.2 & 2.7 & -33.7 & -32.4 & -32.8 &  &  &  &  &  &  &  &  & 3.4 &  2.06 &  0.28 & -0.29 & \\
G331.60$-$0.14 & 16 12 42.8 & -51 26 42.7 & 2.8 & -96.5 & -93.9 & -95.6 & 7.8 & -97.3 & -94.0 & -96.0 & 2.7 & -96.3 & -94.6 & -95.9 & 9.9 &  4.98 &  0.22 & -0.21 & CS \\
G331.65$-$0.25 & 16 13 28.8 & -51 29 34.6 & 4.4 & -38.3 & -34.8 & -36.5 & 3.5 & -39.1 & -35.4 & -36.5 &  &  &  &  & 5.7 &  2.27 &  0.27 & -0.28 & S \\
G331.70$-$0.03 & 16 12 43.5 & -51 17 20.4 & 3.1 & -26.1 & -25.2 & -25.4 & 1.7 & -25.5 & -24.7 & -25.3 &  &  &  &  & 3.6 &  1.63 &  0.30 & -0.31 & \\
G331.72$+$0.45 & 16 10 43.2 & -50 55 38.7 & 4.2 & 18.9 & 21.1 & 20.0 & 3.6 & 18.2 & 21.3 & 19.8 &  &  &  &  & 4.9 &  6.97 &  0.62 & -0.57 & S \\
G331.76$-$0.33 & 16 14 19.0 & -51 28 22.3 & 28 & -36.1 & -23.9 & -29.1 & 24 & -34.1 & -24.2 & -29.7 & 3.6 & -30.4 & -24.5 & -25.6 & 40 &  1.79 &  0.29 & -0.30 & W \\
G331.95$+$0.08 & 16 13 23.5 & -51 02 50.8 & 2.8 & -102.8 & -98.4 & -101.9 & 2.5 & -99.7 & -102.5 & -99.7 &  &  &  &  & 4.2 &  5.19 &  0.22 & -0.21 & \\
G331.96$-$0.24 & 16 14 52.6 & -51 16 07.2 & 3.0 & -81.5 & -80.2 & -80.8 &  &  &  &  &  &  &  &  & 3.9 &  4.35 &  0.21 & -0.21 & W \\
G331.99$-$0.04 & 16 14 05.7 & -51 06 12.7 & 2.9 & -99.1 & -97.1 & -98.0 & 2.8 & -98.2 & -98.6 & -98.2 &  &  &  &  & 3.9 &  5.07 &  0.21 & -0.21 & \\
G332.14$-$0.51 & 16 16 52.0 & -51 20 23.1 & 5.1 & -36.7 & -35.2 & -36.5 & 6.9 & -36.9 & -35.4 & -36.3 &  &  &  &  & 4.5 &  2.28 &  0.27 & -0.28 & \\
G332.22$+$0.36 & 16 13 25.5 & -50 39 30.7 & 3.4 & -121.2 & -120.6 & -121.0 & 2.1 & -122.2 & -122.7 & -122.2 &  &  &  &  & 4.4 &  6.17 &  0.32 & -0.27 & \\
G332.30$+$0.44 & 16 13 27.9 & -50 32 16.3 & 6.2 & -56.3 & -54.1 & -55.4 & 3.6 & -56.4 & -54.4 & -55.3 &  &  &  &  & 8.4 &  3.24 &  0.23 & -0.24 & \\
G332.40$-$0.33 & 16 17 16.5 & -51 01 21.2 &  &  &  &  & 2.7 & -7.4 & -5.0 & -5.6 &  &  &  &  & 3.8 &  4.61 &  0.41 & -0.38 & \\
G332.83$+$0.19 & 16 16 57.1 & -50 21 10.2 & 4.8 & -34.3 & -31.7 & -33.7 &  &  &  &  &  &  &  &  & 5.1 &  2.15 &  0.27 & -0.29 & \\
G332.90$-$0.37 & 16 19 41.3 & -50 42 31.2 & 2.2 & -102.1 & -99.7 & -101.0 & 3.4 & -101.7 & -102.5 & -101.7 &  &  &  &  & 4.9 &  5.20 &  0.21 & -0.20 & \\
G333.06$-$0.10 & 16 19 14.4 & -50 24 16.6 & 4.4 & -48.9 & -43.3 & -45.4 & 6.8 & -48.5 & -43.7 & -45.7 &  &  &  &  & 9.5 &  2.79 &  0.25 & -0.26 & W \\
G333.44$-$0.07 & 16 20 49.1 & -50 06 52.0 & 3.3 & -96.9 & -95.4 & -96.3 & 2.7 & -97.1 & -98.6 & -97.1 &  &  &  &  & 4.2 &  5.00 &  0.20 & -0.20 & \\
G333.46$-$0.34 & 16 22 06.3 & -50 17 29.4 & 4.3 & -63.0 & -56.7 & -58.5 & 4.3 & -64.7 & -58.6 & -64.3 &  &  &  &  & 5.4 &  3.55 &  0.22 & -0.23 & \\
G333.53$+$0.41 & 16 19 06.9 & -49 42 21.9 & 3.3 & -145.1 & -144.7 & -144.9 &  &  &  &  &  &  &  &  & 4.0 &  7.52 &  0.77 & -0.77 & \\
G333.60$-$0.49 & 16 23 21.6 & -50 18 09.7 & 4.1 & -16.1 & -15.0 & -15.7 & 3.0 & -16.3 & -15.7 & -15.9 &  &  &  &  & 4.8 &  1.06 &  0.34 & -0.36 & \\
G333.64$-$0.37 & 16 22 59.3 & -50 11 06.3 & 4.1 & -70.6 & -67.8 & -69.5 & 4.5 & -70.8 & -69.1 & -70.2 &  &  &  &  & 6.3 &  3.93 &  0.21 & -0.21 & W \\
G333.72$-$0.39 & 16 23 25.8 & -50 08 36.6 & 5.2 & -44.3 & -40.4 & -43.3 & 5.0 & -45.7 & -41.5 & -43.5 &  &  &  &  & 7.1 &  2.71 &  0.25 & -0.26 & \\
G333.78$-$0.36 & 16 23 35.6 & -50 04 53.2 & 2.1 & -93.4 & -92.1 & -92.4 & 1.6 & -93.1 & -92.7 & -92.9 &  &  &  &  & 3.1 &  4.85 &  0.20 & -0.20 & \\
G333.90$-$0.09 & 16 22 56.6 & -49 48 07.4 & 3.8 & -126.0 & -123.2 & -123.9 & 4.0 & -123.8 & -122.9 & -123.1 &  &  &  &  & 4.5 &  6.12 &  0.26 & -0.23 & S \\
G334.01$-$0.02 & 16 23 06.3 & -49 40 30.9 & 3.3 & -94.3 & -93.0 & -93.4 & 3.3 & -95.5 & -93.6 & -93.8 &  &  &  &  & 3.9 &  4.89 &  0.19 & -0.19 & \\
G334.11$-$0.37 & 16 25 05.0 & -49 50 58.6 & 3.9 & -32.8 & -30.9 & -31.5 &  &  &  &  &  &  &  &  & 4.7 &  2.08 &  0.28 & -0.30 & \\
G334.20$+$0.22 & 16 22 49.6 & -49 22 18.8 & 2.7 & -7.4 & -7.0 & -7.2 & 2.8 & -5.2 & -7.4 & -5.2 &  &  &  &  & 4.4 &  4.78 &  0.42 & -0.39 & \\
G334.35$-$0.36 & 16 26 03.3 & -49 40 15.0 & 3.2 & -22.8 & -21.5 & -22.6 &  &  &  &  &  &  &  &  & 4.1 &  1.54 &  0.31 & -0.33 & \\
G334.35$+$0.06 & 16 24 13.2 & -49 22 24.8 & 3.8 & -82.6 & -81.9 & -82.1 &  &  &  &  &  &  &  &  & 3.8 &  4.44 &  0.20 & -0.20 & \\
G334.41$+$0.22 & 16 23 44.6 & -49 13 15.1 & 3.9 & -32.8 & -31.7 & -32.2 & 3.0 & -32.5 & -32.1 & -32.3 &  &  &  &  & 4.7 &  2.13 &  0.28 & -0.30 & \\
G334.49$-$0.37 & 16 26 41.1 & -49 34 40.5 & 3.8 & -9.8 & -8.5 & -9.1 & 2.7 & -10.2 & -8.9 & -10.0 &  &  &  &  & 4.7 &  0.62 &  0.38 & -0.41 & \\
G334.61$-$0.43 & 16 27 29.9 & -49 32 14.1 & 5.0 & -139.5 & -138.2 & -138.8 & 3.3 & -139.7 & -138.9 & -139.3 &  &  &  &  & 5.7 &  6.97 &  0.36 & -0.36 & \\
G334.68$+$0.44 & 16 23 57.9 & -48 52 53.8 & 3.6 & -96.9 & -95.2 & -95.2 &  &  &  &  &  &  &  &  & 3.8 &  4.96 &  0.19 & -0.19 & \\
G334.68$+$0.35 & 16 24 22.8 & -48 56 34.8 & 4.0 & -55.6 & -55.2 & -55.2 & 2.6 & -55.9 & -56.4 & -55.9 &  &  &  &  & 4.4 &  3.34 &  0.23 & -0.23 & \\
G334.68$+$0.13 & 16 25 20.5 & -49 05 30.2 & 5.1 & -94.5 & -89.1 & -91.1 & 2.2 & -93.4 & -91.8 & -92.5 &  &  &  &  & 6.1 &  4.83 &  0.19 & -0.19 & W \\
    \hline
  \end{tabular}
  \label{tab:sio}
  \flushleft\small
  $^1$C - presence of a 1665\,MHz \oh maser from \citet{caswell98}; S - presence of a 1612\,MHz \oh maser from \citet{sevenster97}; W - presence of a \water maser from \citet{walsh11,walsh14}.
\end{table*}
\end{landscape}
\twocolumn
%--------------------

%--------------------
\twocolumn[{
    \begin{minipage}{\textwidth}
      \includegraphics[width=1.0\textwidth]{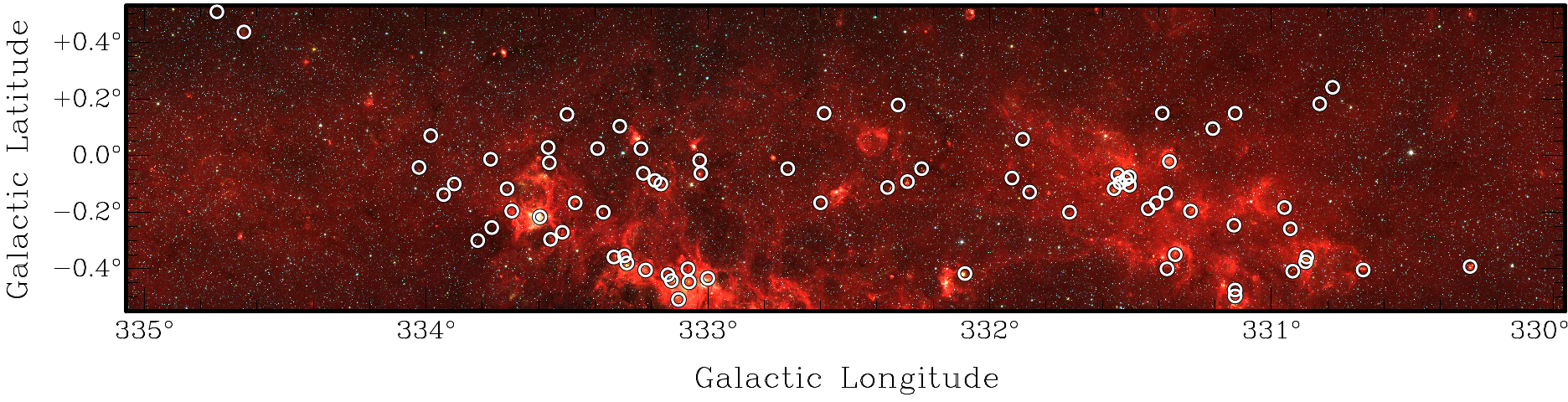}
      \captionof{figure}{{\em Spitzer} GLIMPSE 3-colour (RGB = 3.6, 4.5 and 8.0\,$\mu$m) image with MALT-45 positions of class I \choh masers (circle symbols).}
      \label{fig:ch3oh_glimpse}
%--------------------
      \vspace{5mm}
      \includegraphics[width=1.0\textwidth]{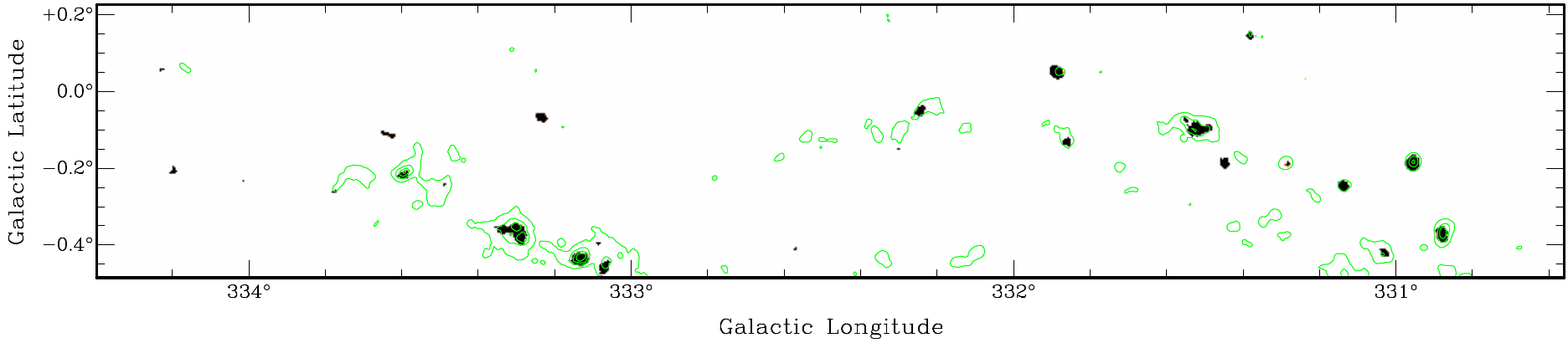}
      \captionof{figure}{Auto-correlated map of \sio (1--0) $v=0$, overlaid with \cs (1--0) thermal emission contours. This image shows \sio emission at and above 0.04\,K ($\sim$1.5$\sigma$). Contour levels are 14, 34, ...74 per cent of 2.62\,K.}
      \label{fig:cs_sio}
%--------------------
      \vspace{5mm}
      \includegraphics[width=1.0\textwidth]{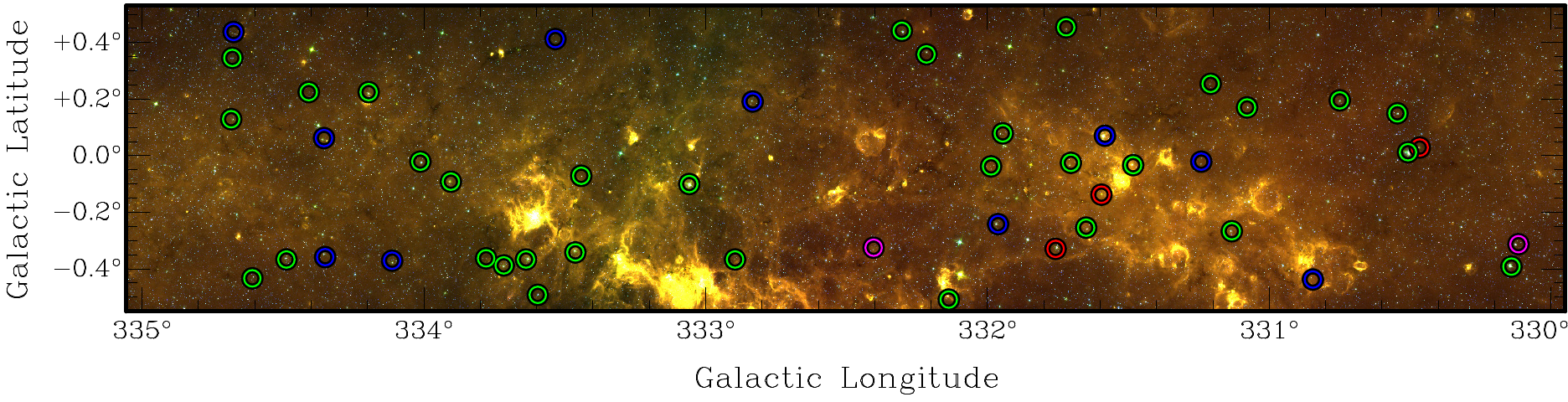}
      \captionof{figure}{{\em Spitzer} GLIMPSE 3-colour (RGB = 3.6, 5.8 and 8.0\,$\mu$m) image with MALT-45 positions of detected \sio (1--0) $v=1,2,3$ masers (circle symbols). The circles have been colour-coded by the vibrational modes within each region. Red: region containing $v=1,2,3$ maser emission; green: $v=1,2$; blue: $v=1$; magenta: $v=2$.}
      \label{fig:sio_glimpse}
      \vspace{10mm}
    \end{minipage}
}]
%--------------------

The positions of class I \choh masers have been compared with the published positions of class II \choh, \water and \oh masers to establish associations. When considering the maximum distance between masers that still constitute an association, we use 60 arcsec from the class I position as suitable for defining associations. \citet{csengeri14} analysed a large distribution of clumps associated with HMSF in their source sample, and find that all clumps have effective radii within 55 arcsec. Additionally, we note that increasing the maximum offset beyond 60 arcsec does not increase the number of unique associations. These findings suggest for our purposes that 60 arcsec is sufficient to group masers within a single star-forming region. We have chosen not to define an association based on source distance or size, as it is difficult to implement across a wide range of objects, kinematic distances may be erroneous, and maser geometry may be wide-spread. An angle of 60 arcsec on the sky corresponds to approximately 1.5\,pc at a distance of 5\,kpc.

There are two targets where the spread of class I \choh maser emission is large and these have been considered as two separate maser regions. The weaker masers (G331.52$-$0.08 and G333.14$-$0.42) near to each of these bright masers (G331.50$-$0.08 and G333.13$-$0.44) possess a distinct spectral velocity component and are sufficiently spatially offset to be considered a different site (54 and 87 arcsec, respectively). See the spectra in Appendix B (available online). These sources are marked with a ``P'' in Table \ref{tab:ch3oh}.

Integrated intensity values for \cs and \sio $v=0$ have been included for each maser, where possible, in Table \ref{tab:ch3oh}. Emission was integrated $\pm$10\,\kms from the peak velocity of the maser. Values were not included if the emission was not significant ($<3\sigma$).

\subsection{\sio (1--0) $v=0$ thermal emission at 43.424\,GHz}
Extended \sio $v=0$ emission is detected in some regions, as shown in Figure \ref{fig:cs_sio}, which shows that \sio $v=0$ emission tends to be associated with bright peaks of \cs emission. Note that emission from the \sio $v=0$ is typically weaker than for \cs, hence the signal-to-noise is lower for this transition; Figure \ref{fig:cs_sio} shows all emission above $1.5\sigma$. Discussion on the relation between class I \choh masers and \sio $v=0$ emission is contained in Section \ref{sec:sio_ch3oh}.

\subsection{\sio (1--0) $v=1,2,3$ masers at 43.122, 42.820 and 42.519\,GHz}
The vibrationally excited \sio (1--0) lines were detected towards \siototal regions, as shown in Figure \ref{fig:sio_glimpse}. The properties of each maser region are classified in Table \ref{tab:sio}. The search procedure for \sio masers was the same as with \choh masers; they were identified by visual inspection of the data, and had velocities categorised by a $1.5\sigma$ cutoff from the peak velocity. Velocities are specified to 0.1\,\kms, and flux densities to two significant figures. Many masers have low signal-to-noise values, but are considered real detections based on their association in GLIMPSE, and simultaneous detection of emission in another \sio maser transition. The positions of \sio masers have been compared with \oh and \water masers; see Table \ref{tab:sio} and Section \ref{sec:sio_discussion}. \choh maser species were searched at these positions, but none were found, implying that any class I \choh maser emission associated with the \sio masers has a peak flux density $<$4.5\,Jy. Peak intensity maps are presented in Appendix C, and each un-smoothed maser spectrum can be seen in Appendix D. Appendices are available online.

\begin{figure}
  \includegraphics[width=0.47\textwidth]{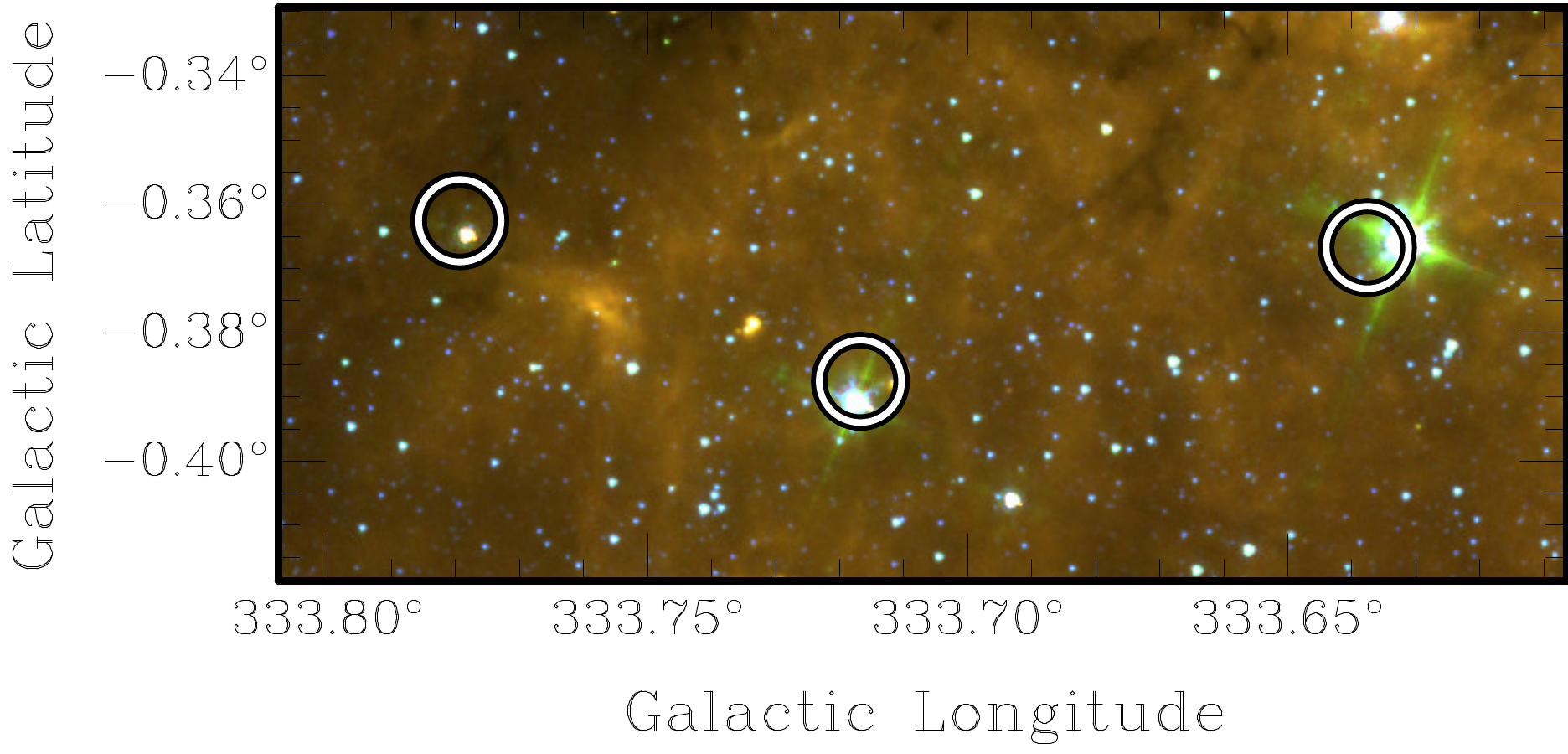}
  \caption{{\em Spitzer} GLIMPSE 3-colour (RGB = 3.6, 5.8 and 8.0\,$\mu$m) image with examples of detected \sio masers towards infrared stars. Circles are centred on the peak position of maser emission, and are beam-sized (1 arcmin). All three of these \sio maser regions contain the $v=1$ and $v=2$ lines, but all other maser region variants have similar infrared colours.}
  \label{fig:sio_glimpse_example}
\end{figure}

Figure \ref{fig:sio_glimpse} shows the locations of \sio masers detected by MALT-45 with respect to infrared emission from GLIMPSE. All \sio masers are detected towards point-like, isolated sources that appear to be evolved stars, based on their infrared colours and lack of obvious associations with star formation, such as IRDCs or extended infrared emission. Figure \ref{fig:sio_glimpse_example} shows the relative locations of three \sio masers compared to infrared stars; these masers were chosen as they show various colours of infrared stars towards all \sio masers. It can be seen that the maser locations are not identical to the stars, but are within one beam. Therefore, we cannot conclusively argue that each \sio maser site is coincident with an evolved star, although the evidence we have suggests this is the case. Further higher spatial resolution observations of the maser sites will be able to demonstrate conclusively whether or not each \sio maser is associated with an evolved star.

Of the 47 \sio maser regions, 45 contain $v=1$ emission, 37 contain $v=2$, and only 3 have $v=3$ emission. We have found no previously published data for any of these \sio masers and consider them all to be new detections. Some \sio maser regions have greater $v=2$ emission than $v=1$, and two have no detected emission $v=1$; this is discussed further in Section \ref{sec:sio_discussion}. There are also 12 \sio maser regions with an associated \oh or \water maser; see Section \ref{sec:sio_discussion}.

Parameters used in determining kinematic distances are the same as those found in Section \ref{sec:results_ch3oh}, except for a more generous velocity uncertainty $\sigma(V_{LSR})=5.0$\,\kms, and $V_s=0$\,\kms. Four masers are labelled at the far kinematic distance, because the position and errors for a near distance are unrealistic; see Table \ref{tab:sio}.

\subsection{Survey completeness}
By analysing the RMS noise values per pixel distributed across the MALT-45 region, we find that the noise levels may vary by up to a factor of three due to some data being flagged or insufficient time on the telescope to complete the observations. However, we determine that approximately 95 per cent are consistent with the values listed in Table \ref{tab:spectral_lines}, and only 5 per cent show RMS levels worse than these values. Thus, we assess that to first order the survey has uniform sensitivity and only a small fraction of the area has sensitivity poorer than nominal.

Tables \ref{tab:ch3oh} and \ref{tab:sio} include the RMS value for each detected maser peak, which help to identify the completeness of the MALT-45 survey. We detect many masers at less than $5\sigma$ significance. However, we consider the survey complete at the $5\sigma$ level for 95 per cent of the survey area. This completeness limit is equivalent to 4.5\,Jy for class I \choh masers, and approximately 4.3\,Jy for each \sio maser line.

%%%%%%%%%%%%%%%%%%%%%%%%%%%%%%%%%%%%%%%%%%%%%%%%%%%%%%%%%%%%%%%%%%%%%%%%%%%%%%%%
\section{Discussion}
\subsection{Comparing \cs (1--0) with HOPS \nh (1,1)}
\label{sec:cs_nh3}
\begin{figure*}
  \includegraphics[width=1.0\textwidth]{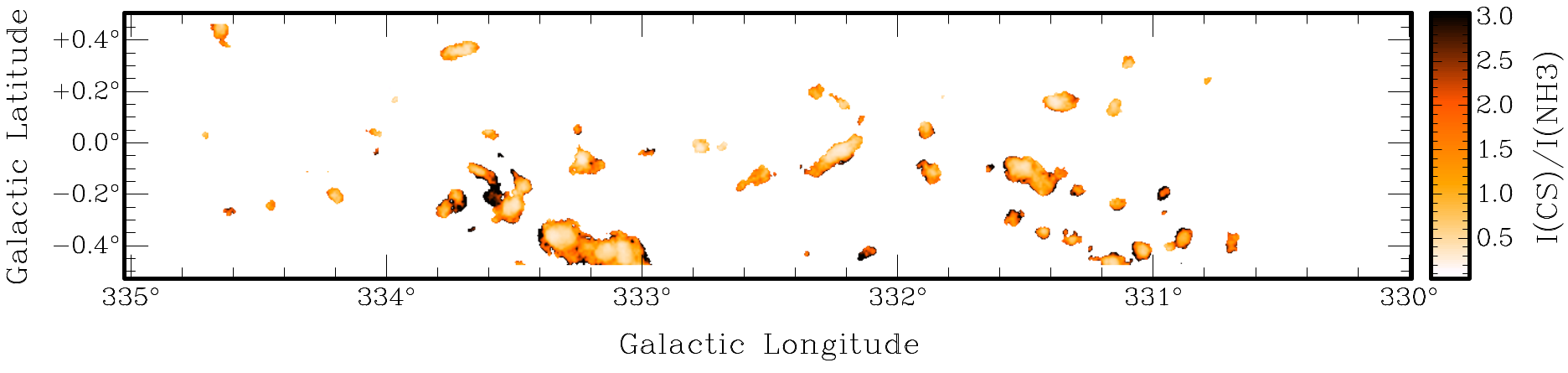}
  \caption{A ratio image of integrated MALT-45 \cs (1--0) to HOPS \nh (1,1). The clumps visible here contain at least 0.5\,K\,\kms of integrated \nh (1,1) emission. Most clumps have edge values around 2, and centres of clumps can have values as low as 0.4.}
  \label{fig:ratio}
\end{figure*}

Figure \ref{fig:cs_hops} shows that \nh (1,1) from HOPS is closely correlated with bright \cs (1--0) emission, but with prominent exceptions; examples include G330.30$-$0.40, G332.17$-$0.10 and G334.17$+$0.06. We calculate that MALT-45 is more sensitive to \cs (1--0) emission than HOPS is to \nh (1,1) by a factor approximately 9, therefore \cs without \nh in Figure \ref{fig:cs_hops} may be due to the superior sensitivity of MALT-45 over HOPS. However, it is also likely attributable to the phenomena described by \citet{taylor98}, who reason that \cs is an ``early-type'' star-forming molecule, tracing quiescent gas, while \nh is ``late-type'', associated with dense cores. For this reason, despite having a higher effective critical density, \cs without \nh may be tracing molecular clouds before cold, dense cores have formed. Additionally, \citet{bergin01} and \citet{tafalla02} indicate that \cs tends to deplete in core collapse. This means a relatively high abundance of \nh to \cs should reflect a cold, dense core, which is a \citet{taylor98} ``late-type'' scenario. As results indicate later in this section, \nh is not persistent in even further developed stages of HMSF, as it is not detected or very faint. We distinguish ``late-type'' sources from these evolved sources through bolometric luminosities and the presences of infrared features, such as stars or \hii regions.

We produced a map of the ratio between the integrated intensities of \cs and \nh in order to highlight the differences between their distributions. Before performing the quotient of integrated intensity, the \cs data were smoothed to the HOPS \nh resolution ($\sim$2\,arcmin). We used a cutoff at 0.5\,K\kms ($3\sigma$) of the less sensitive \nh data; \cs data is $>3\sigma$ everywhere \nh is $>3\sigma$. Integration of emission was taken between -100 and -40\,\kms, to eliminate subtle baseline effects which decrease the robustness of analysis; this velocity range is where most of the emission in this part of the Galaxy lies (see Figure \ref{fig:cs_thrumms} for the longitude-velocity plot). Quotients use units of main-beam temperature for both \cs and \nh. See Figure \ref{fig:ratio} for the full ratio map.

Moving inward from the clump edges, the \cs/\nh ratio quickly changes. We observe the amount of both \cs and \nh to increase, with the \nh intensity increasing more ``quickly''. Edge values are mostly about 2, while clump centres are usually between 1 and 0.5, but can be as low as 0.4.

We now discuss the ratio map qualitatively, reserving radiative transfer modelling for future work. In almost every clump, the centre is dominated by \nh over \cs. This could be due to any or a combination of enhanced \nh, depleted \cs, or both species being enhanced but \cs becoming optically thick. The hyperfine line structure of \nh means that it remains optically thin at higher column densities than \cs. To test for optically thick \cs, we compared \cs with MALT-45 C$^{34}$S.

We conducted the analysis of \cs against C$^{34}$S in a similar manner to that of \cs against \nh. Both integrated maps were smoothed to the same resolution, the velocity range of integration was -100 to -40\,\kms, and emission was limited to detectable C$^{34}$S (0.15\,K\,\kms). A preliminary analysis of the C$^{34}$S data was undertaken only to investigate the \cs optical depth within these clumps.

\begin{figure}
  \subfigure{
    \includegraphics[width=0.47\textwidth]{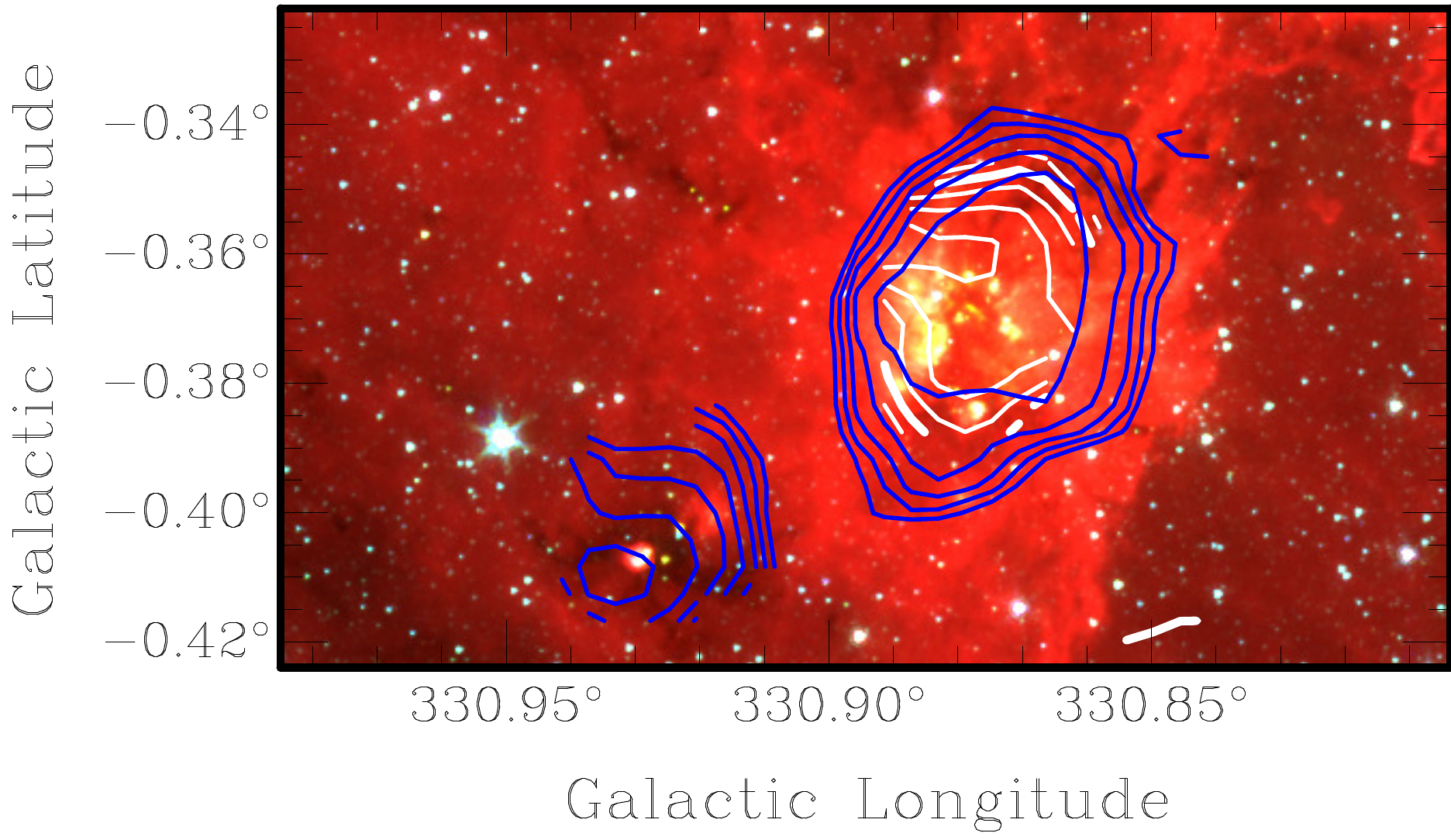}
  }
  \subfigure{
    \includegraphics[width=0.47\textwidth]{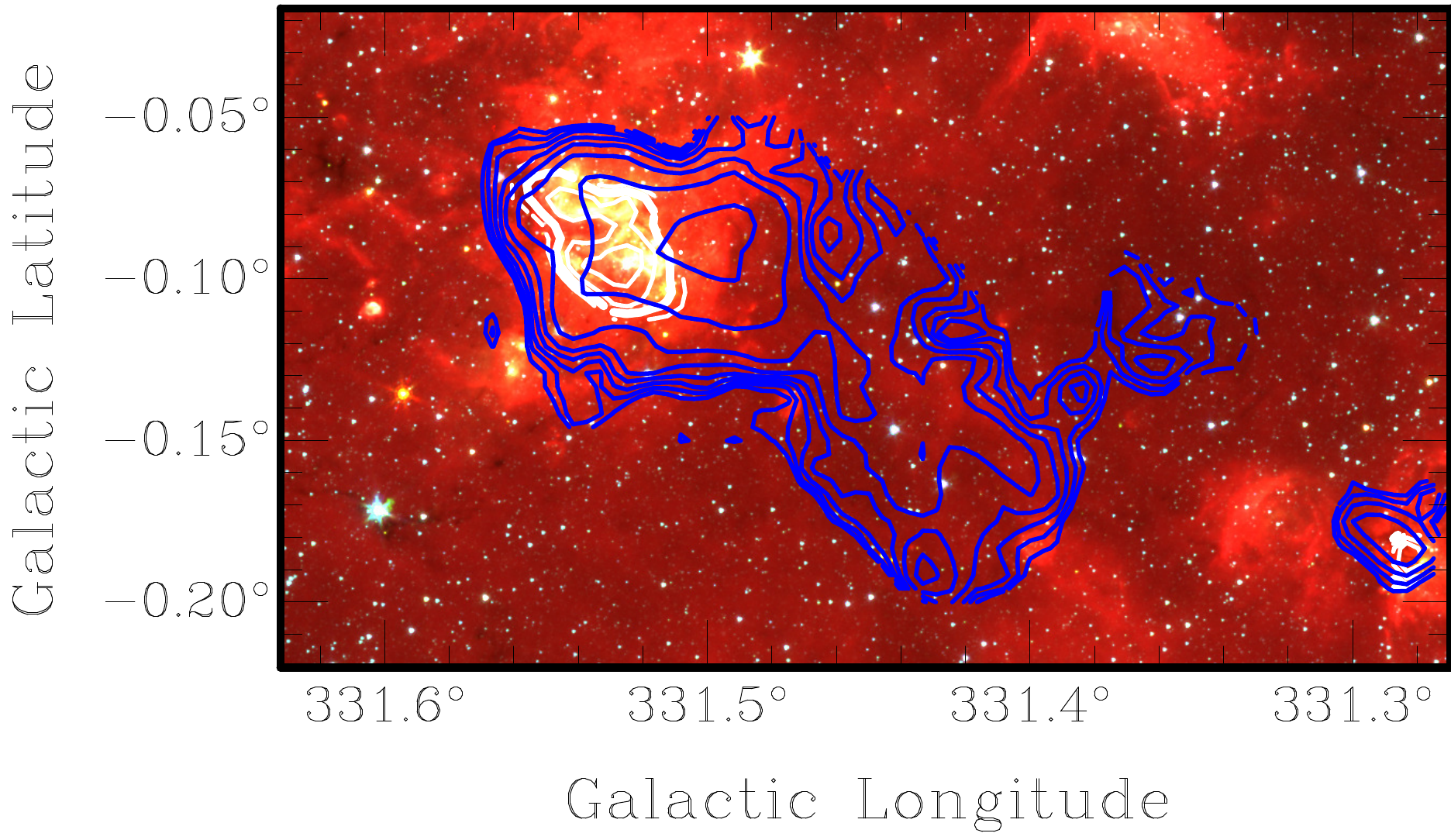}
  }
  \subfigure{
    \includegraphics[width=0.47\textwidth]{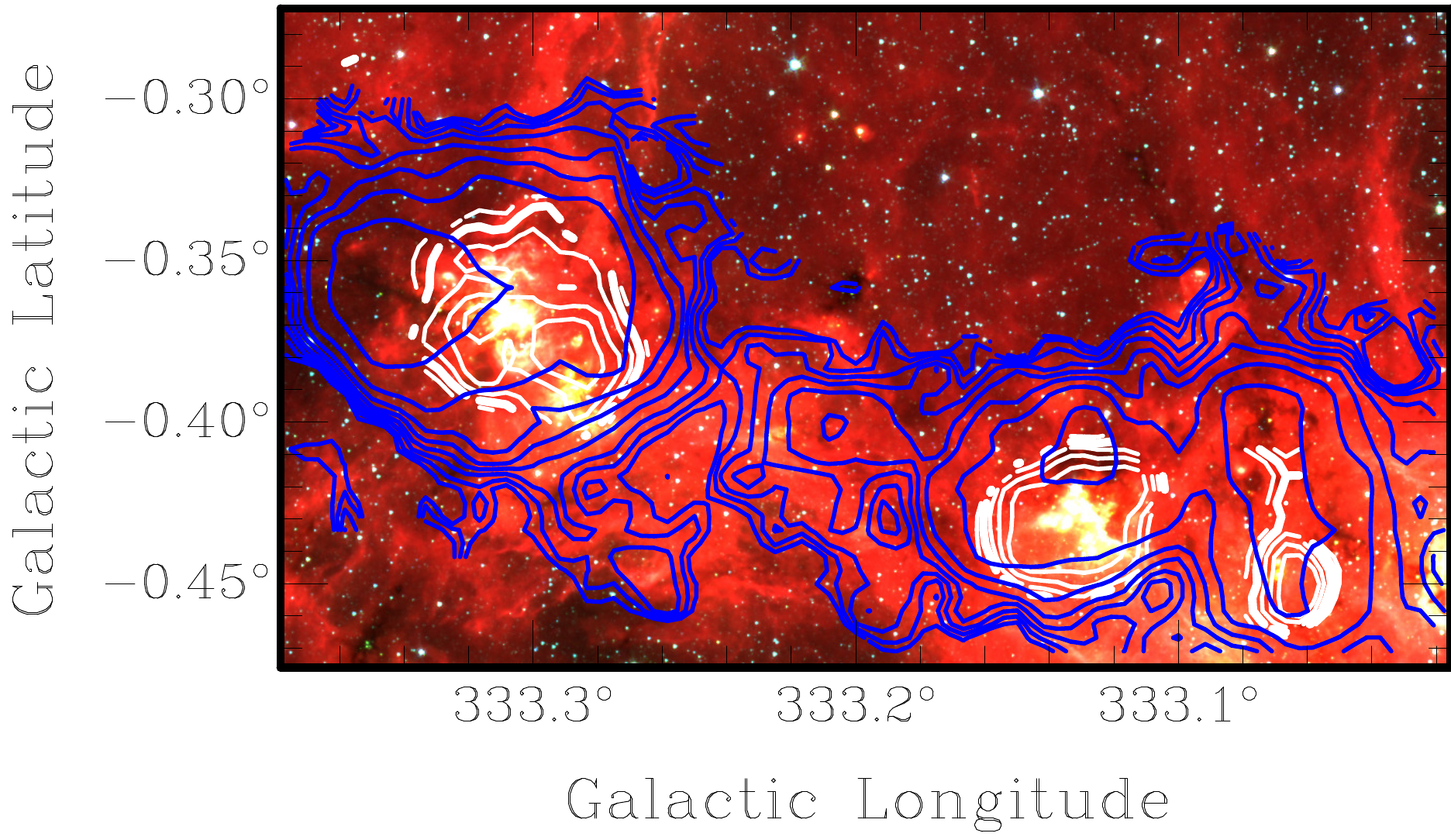}
  }
  \caption{Examples of significant C$^{34}$S/\cs ratios within \cs/\nh clumps. {\em Spitzer} GLIMPSE 3-colour (RGB = 3.6, 4.5 and 8.0\,$\mu$m) image with contours of I(C$^{34}$S)/I(\cs) (white) and contours of I(\cs)/I(\nh) (blue). I(C$^{34}$S)/I(\cs) contour levels are 0.08, 0.10, ...0.16, and I(\cs)/I(\nh) contour levels are 0.3, 0.5, ...1.9. Note that contour levels for I(\cs)/I(\nh) move from large to small, outside-in. The lowest \cs/\nh and highest C$^{34}$S/\cs ratios appear to be slightly offset. High C$^{34}$S/\cs ratios appear to occur with bright infrared emission; this is likely due to C$^{34}$S is only detectable where the overall \cs abundance is enhanced in hotter, shocked regions where these molecules have been released into the gas phase.}
  \label{fig:c34s_glimpse}
\end{figure}

Figure \ref{fig:c34s_glimpse} shows contours of the ratio of C$^{34}$S/\cs against GLIMPSE, with contours of \cs/\nh. Progressively higher levels of C$^{34}$S/\cs indicate \cs self-absorption, but appear to be slightly offset from the lowest \cs/\nh ratios. This implies that the relatively fainter \cs emission in the presence of an abundance of \nh is due to both depletion and optical thickness. Additionally, the higher ratios of C$^{34}$S/\cs appear to be associated with bright infrared emission, and correlate well thermal \sio emission. We speculate that high C$^{34}$S/\cs ratios may be due to high \cs optical depth as a result of high \cs relative abundance in the presence of an outflow or shocked gas. C$^{34}$S is not detected in other regions of these clumps; aside from having poor signal-to-noise, this is likely due to the isotopologue depleting similar to \cs, and emission is only apparent in hotter, shocked regions where these molecules have been released into the gas phase.

A similar analysis to the one we have conducted was performed in \citet{tafalla04} towards two cold, dense cores. Their results indicate significant depletion of \cs and C$^{34}$S towards the core centres, while \nh peaks in abundance. We have not undertaken a detailed analysis of the temperature or mass of the MALT-45 molecular gas clumps in this paper; this will be investigated in a future publication focusing on the \cs emission. In future work, we will present radiative transfer modelling (including depletion, time-dependant chemistry and the relative sensitivity limits of the various surveys) of the dust, \cs and \nh emission to quantitatively understand the physical reason for these effects.

The following subsections briefly discuss three unusual clumps, which have the largest \cs/\nh ratios.

\subsubsection{G330.95$-$0.19}
This source is one of the brightest in \cs across the MALT-45 survey region (peak of 1.5\,K), and is the brightest in thermal \sio emission (peak of 0.17\,K). The source also contains all the common star formation region masers: class I and II \choh, \water and \oh. The \cs spectrum has a broad profile of $\sim$10\,\kms FWHM, possibly due to a powerful molecular outflow. A strong outflow would not be surprising given the strength of the thermal \sio detected, relative to all other \sio sources in MALT-45.

\citet{garay10} have targeted this source which has a high IRAS far-infrared luminosity, as well as detection of \cs (2--1) by \citet{bronfman96}. \citet{garay10} present data for \sio (2--1) and note that of the four sources they discuss, this source is brightest in \sio, and conclude that an outflow is present due to the spectral profiles of \cs and \sio. \citet{garay10} also discuss the requirement of OB stars to produce the ionised gas seen in other observations. The evolved nature of the source is the likely reason for the faint \nh, because it is being destroyed or dispersed by outflows or young stellar objects.

\subsubsection{G332.10$-$0.42}
The \cs/\nh ratio of this source is approximately 3. Of the three unusual sources discussed in this section, this is the only one without detected thermal \sio emission. A small region ($\sim$30 arcsec) of the clump has a constant C$^{34}$S/\cs ratio of 0.08, and is coincident with a \water and class II \choh maser, as well as what appears to be a star in GLIMPSE. A new class I \choh maser is detected by MALT-45, but is slightly offset from this position ($\sim$30 arcsec).

This source has been targeted by \citet{ilee13} to determine disc properties through \co observations, and is listed with a very high bolometric luminosity ($1.8\times10^5$ L$_\odot$). The Red MSX Source (RMS; \citealt{lumsden13}) survey also lists a {\em K}-band magnitude of 5.9. These values suggest that the source is not embedded. As a star is visible in GLIMPSE and the source is not embedded, it may be assumed that the star (and the region) are post star-forming. However, we are confident that the associated star is not evolved (beyond the main-sequence), as a near-infrared (NIR) spectrum of the source contains Br\,$\gamma$, indicating the presence of an \hii region. The spectrum is taken from the RMS survey, although the data remain unpublished; \citet{cooper13} present detailed analysis of the NIR spectroscopy, but note that it does not cover this specific source. We conclude that this source is somewhat evolved because of its exposure from natal clouds, but is pre-main-sequence because of the class II \choh maser emission. Again, the evolved nature of this region is the likely reason for the relatively faint \nh.

\subsubsection{G333.60$-$0.20}
This region is the brightest \hii region within the MALT-45 survey, and appears to harbour evolved sources. C$^{34}$S emission is detected within this source, and the C$^{34}$S/\cs ratio varies from 0.09 to 0.14. Thermal \sio emission is also detected towards the source. As well as a previously known class I \choh maser, this region is associated with 1612-, 1665- and 1667-MHz \oh masers \citep{sevenster97,caswell98}, and many \water masers from HOPS \citep{walsh14}. The peak velocity of the class I \choh maser is $-49.7$\,\kms, $-51$\,\kms for \oh, and $-75.5$ to $-43.1$\,\kms for the \water masers. MALT-45 \cs also has a double-peak profile in this region; a strong, broad peak is associated with $-48.1$\,\kms and a weaker one at $-89.8$\,\kms, both at the same position (16$^h$22$^m$06$^s$, $-$50$^\circ$06$^\prime$17$^{\prime\prime}$ J2000).

\citet{fujiyoshi06} conducted observations of H90\,$\alpha$ towards this source, which also reveal two peaks of emission, and reason that it contains many O-type stars. The peak velocity of each H90\,$\alpha$ component varies greatly over the region, from $-78.5$ to $-52.9$\,\kms and $-42.8$ to $-22.8$\,\kms, where the second peak is weaker and not present across the observed area. Note that the position of peak MALT-45 \cs is outside of the data presented in \citet{fujiyoshi06}; it is possible that the spatial resolution of our \cs data is limiting our ability to interpret the physics of this region. However, it is interesting that the peak velocities of \cs are quite different to H90\,$\alpha$; this may indicate complex structure of the source, and indeed this is discussed in \citet{fujiyoshi06}.

The weak presence of \nh (1,1) emission is mysterious. The high \cs/\nh ratio may be attributable to the region being at a higher temperature, preventing \cs depletion. \citet{lowe14} attempted a \nh (1,1) and (2,2) analysis of this clump, but were unable to detect significant emission. They suggested that the relatively faint \nh emission could be due to extreme self-absorption. \citet{lo09} comment that in general, \cs (2--1) and HCO$^+$ (1--0) have similar velocity structure, except at this location; the HCO$^+$ has a strong self-absorption feature. However, given that the source is clearly evolved, it is consistent with the other sources discussed in having little \nh emission.

\subsubsection{Summary}
The cause for the high ratios (\cs/\nh $>$ 3) in our analysis appears to stem from evolved regions of HMSF. The other clumps are more difficult to classify - Are they less evolved? Do they have significantly different temperatures or column densities? More analysis is needed, and will be presented in a future MALT-45 paper with \cs clump analysis.

\subsection{Comparing \cs (1--0) with class I \choh maser regions}
\begin{figure}
  \includegraphics[width=0.47\textwidth]{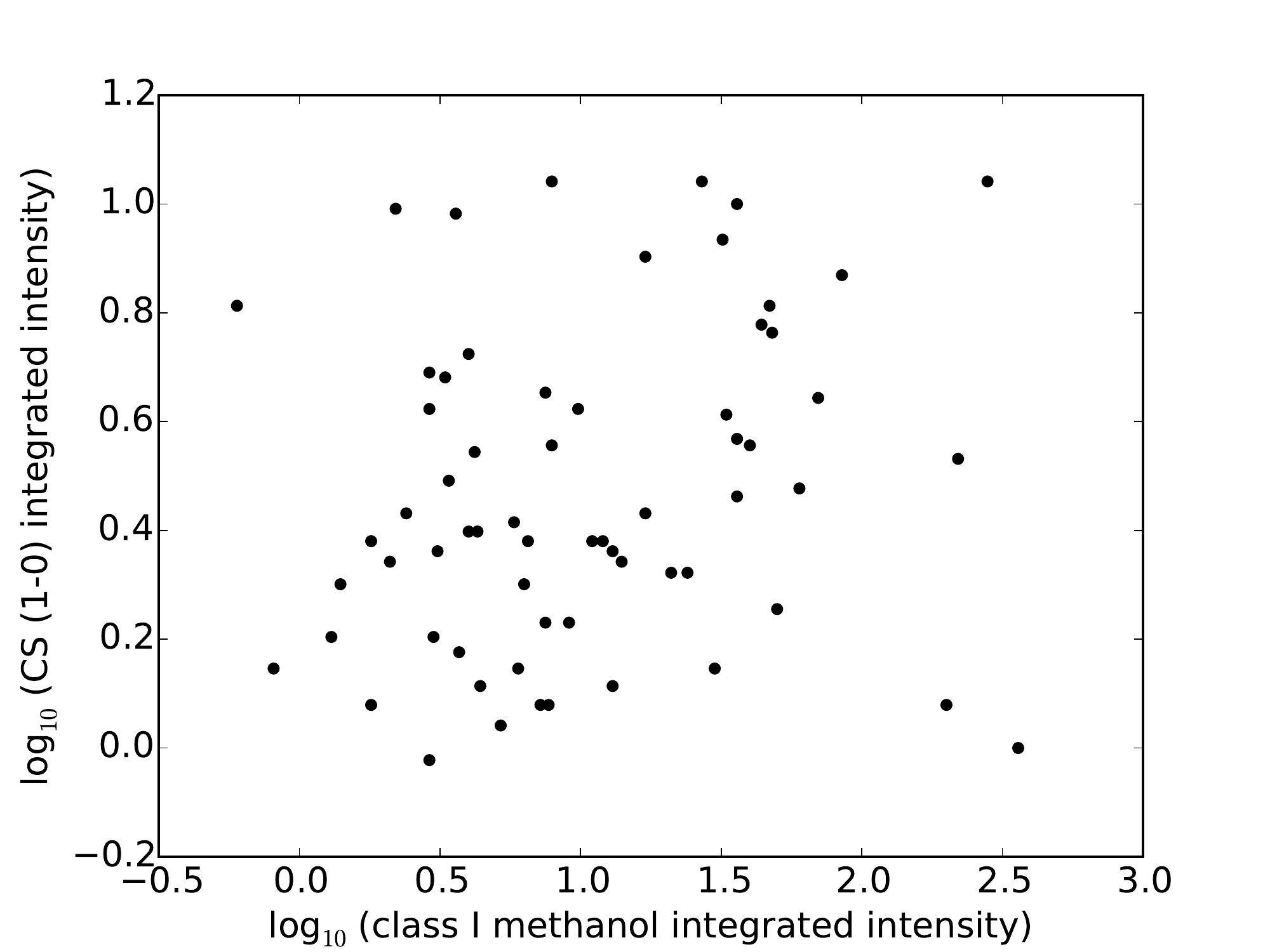}
  \caption{A log-log scatter plot of class I \choh and \cs integrated intensities. Units of integrated intensity are Jy\,\kms for class I \choh, and K\,\kms for \cs. There is no significant correlation observed in either subpopulation, or the total population.}
  \label{fig:cI_cs}
\end{figure}

Table \ref{tab:ch3oh} includes the integrated flux density of \cs (1--0) in each maser region. The kinematic distance calculated for each source is used to determine the luminosities. Figure \ref{fig:cI_cs} compares the integrated \cs intensity with the integrated class I \choh maser intensity.

Figure \ref{fig:cs_ch3oh} shows that most class I \choh masers appear to be associated with a bright peak of \cs emission (73/77 masers with $>$0.11\,K of peak intensity \cs). This is quantified (using integrated intensity) in Figure \ref{fig:cI_cs}. However, there is no correlation observed between the relative intensity of a class I \choh maser and its associated \cs.

A class I \choh maser with \cs emission suggests association with a star-forming region, because sufficient gas and shocks exist to power their emission. However, without being able to associate these lines with other diagnostics of star formation, it is not certain that the masers are stimulated by star formation; cloud-cloud collisions have been found to stimulate class I \choh emission \citep{sjouwerman10}. While the possibility of being a non-star-forming maser extends to all candidates without additional tracers (such as class II \choh masers), we expect that these are exceptional cases, and that the majority if not all class I \choh masers in the MALT-45 sample are associated with star formation.

\begin{figure}
  \includegraphics[width=0.47\textwidth]{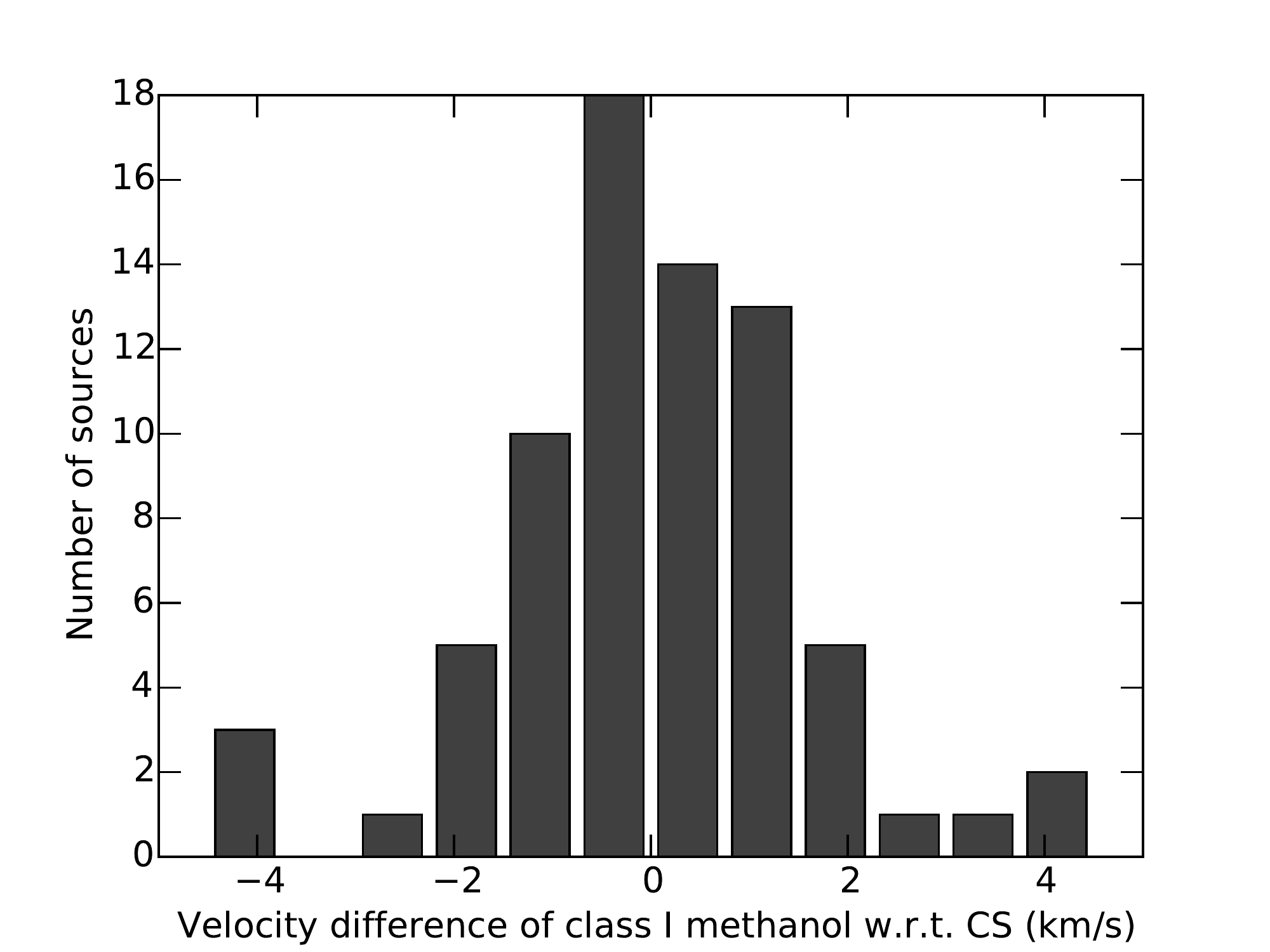}
  \captionof{figure}{The distribution of velocity offsets between the peak velocity of class I \choh and \cs (1--0). The bin size was chosen to be 0.75\,\kms. The histogram is approximated by a Gaussian with a mean velocity 0$\pm$0.2\,\kms and standard deviation 1.5$\pm$0.1\,\kms.}
  \label{fig:cs_ch3oh_vel_hist}
\end{figure}

A comparison of class I and class II \choh masers was undertaken by \citet{voronkov14}, and found a broad Gaussian distribution of velocity differences. We have performed a similar analysis with the peak velocities of both \cs and class I \choh masers from the MALT-45 data. The resulting histogram of the velocity difference of \choh with respect to \cs is found in Figure \ref{fig:cs_ch3oh_vel_hist}. The distribution is approximated by a Gaussian with a mean velocity difference 0$\pm$0.2\,\kms and standard deviation 1.5$\pm$0.1\,\kms. The median velocity difference is -0.1\,\kms.

The FWHM of the population presented by \citet{voronkov14} is approximately 7.8\,\kms, while ours is approximately 3.5\,\kms. Because \cs is a dense gas tracer, it will occur in cloud cores and thus trace systemic motions accurately. Given the resulting Gaussian of velocity differences, this means that class I \choh masers are also good indicators of systemic velocities. \citet{voronkov10a} discuss class I \choh masers having similar velocities to the quiescent gas, due to a small amount of gas actually being shocked in these regions. As is noted in \citet{voronkov14}, \citet{garay02} indicate that \choh emission arising without thermal \sio emission is due to mild shocks, destroying dust grain mantles but not dust grain cores. Hence the dispersion in class I \choh maser velocities is thought to be small, and is reflected here. This also indicates that the large dispersion seen by \citet{voronkov14} is likely due to the large class II \choh maser velocity spread. This is readily explained by class II \choh maser emission radiating from close to a high-mass young stellar object, and therefore having a higher dispersion.

The histogram in Figure \ref{fig:cs_ch3oh_vel_hist} appears to have a slight blueshifted component. This is also discussed in \citet{voronkov14} as perhaps due to maser emission being easier to detect on the near-side of a cloud to the observer. While the median velocity offset is negative, it is statistically insignificant. A larger data set is required to determine if the blueshift phenomena is real or not.

\subsection{Comparing class I \choh maser regions with \sio (1--0)}
\label{sec:sio_ch3oh}
As discussed in Section \ref{sec:spectral_lines}, \sio $v=0$ is a good tracer of shocked gas and outflows. Regions of thermal \sio and positions of class I \choh masers correlate weakly; 23 of 77 masers (30 per cent) have at least a $2\sigma$ peak of \sio emission. As class I masers are collisionally excited, it is perhaps not surprising to find some association with thermal \sio. However, the majority of masers are detected without \sio $v=0$ emission. This could be due to: insufficient sensitivity to extended \sio emission; class I \choh masers having stronger intensity than thermal \sio; or not all class I \choh masers being associated with \sio $v=0$ emission. \citet{garay02} found that thermal \choh emission may be triggered by less energetic shocks, which destroy dust grain mantles but not their cores, which explains a lack of \sio emission. In more energetically shocked regions (such as a jet or molecular outflow), \choh may not survive, which may explain \sio emission without \choh.

Three regions with strong \sio emission (0.04\,K) are not associated with any maser species: G331.02$-$0.42, G333.63$-$0.11 and G334.20$-$0.20. Each of these regions has a \cs peak of at least 0.4\,K ($>10\sigma$), but no obvious infrared emission. Table \ref{tab:ch3oh} lists 28 class I \choh masers with detected \sio emission, implying that these three regions without masers are in the minority. This could be due to the variability of any masers making them undetectable during observations, or that conditions in these regions are unfavourable for masers to exist.

It is clear that sensitivity to \choh and/or \sio will play a part in the detectability of each. However, the fact that we see regions with both \choh and \sio, regions with only \choh masers and regions with only \sio emission suggests that there are multiple conditions under which each of these will exist. If we assume that all class I \choh masers are associated with star formation, then given that the majority of masers are not associated with thermal \sio emission, these masers may be tracing weaker shocks that do not produce detectable \sio emission. 30 per cent of \choh masers also show thermal \sio emission. It is not clear whether the masers and \sio are physically connected, although a reasonable interpretation is that the \sio originates in a jet, where as the masers are associated with the surrounding outflow. In a subsequent paper for MALT-45, maser follow-up observations will be conducted at high spatial resolution. We will simultaneously observe \sio with greater sensitivity, and using this data we will more thoroughly investigate the association of class I \choh masers and thermal \sio emission.

\subsection{Class I \choh maser comparison with GLIMPSE}
\begin{figure}
  \includegraphics[width=0.47\textwidth]{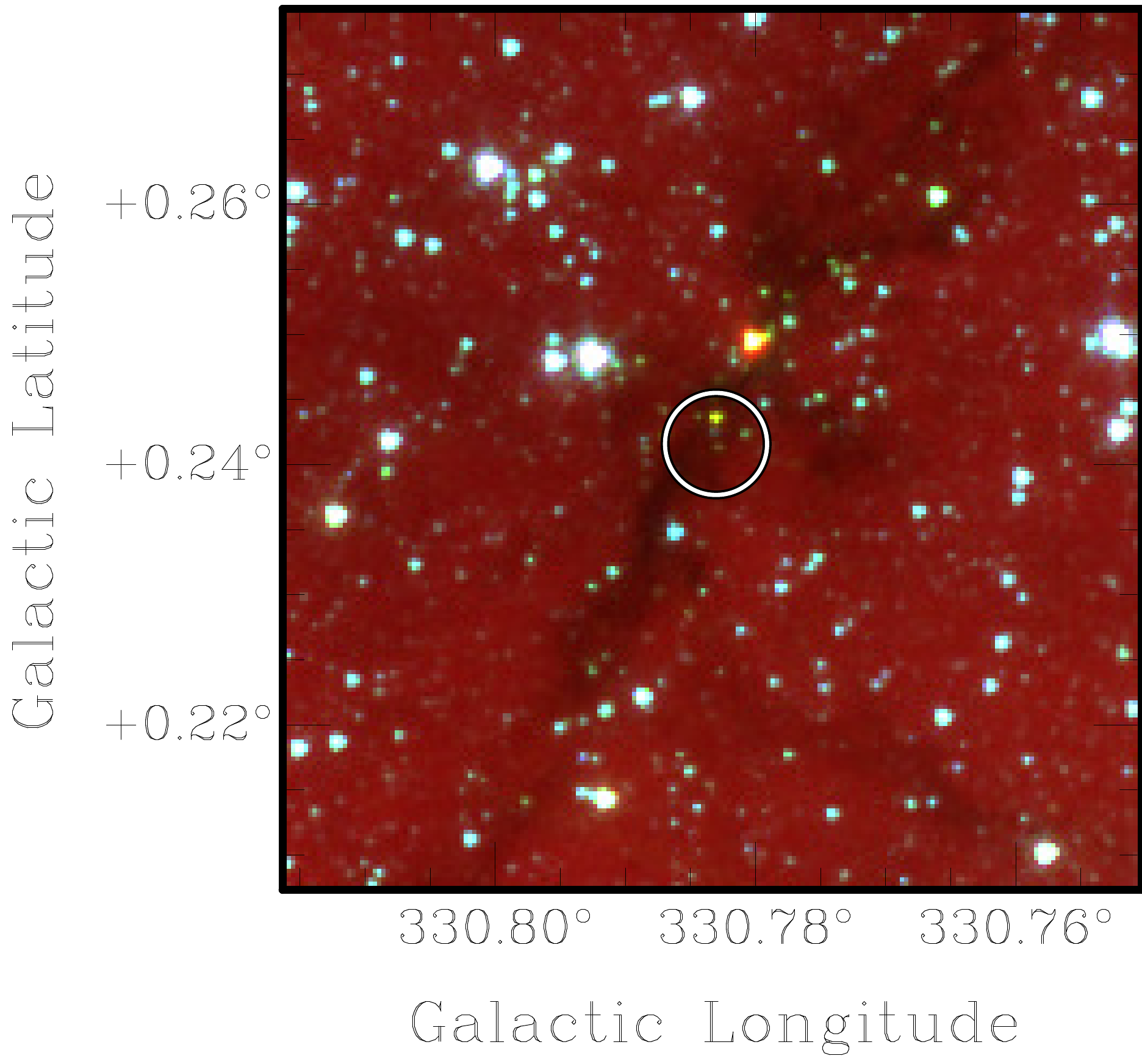}
  \caption{An example of a class I \choh maser with associated ``green'' emission. This region is not classified as an EGO by \citet{cyganowski08}, but may have similar properties. The circle represents 15 arcsec centred on the peak of the class I \choh maser emission, which is the expected positional error.}
  \label{fig:ch3oh_green}
\end{figure}

GLIMPSE mapped the Galactic plane in four infrared bands (3.6, 4.5, 5.8 and 8.0\,$\mu$m). These images are useful for investigating potential relationships between infrared emission and MALT-45 detections. A common example of how GLIMPSE data is used in star formation research is by combining the 3.6, 4.5 and 8.0\,$\mu$m bands in a three-colour image (blue, green and red, respectively) to reveal EGOs (extended green objects). As discussed in \citet{cyganowski08}, EGOs have a strong correlation with class I \choh maser regions.

The catalogue provided by \citet{cyganowski08} was used to investigate the presence of EGOs over the survey region. The catalogue covers $10^{\circ} \leq l \leq 65^{\circ}$ and $295^{\circ} \leq l \leq 350^{\circ}$, $b=\pm1.0^{\circ}$, and lists 22 candidate EGOs within the MALT-45 survey region. Of the 77 class I masers found by MALT-45, 12 are associated with an EGO. Association is credited for any spatial overlap within 60 arcsec. Of the 12 EGO-associated class I \choh masers, 11 were previously known (G330.95$-$0.18 is new). EGOs are classified by visual inspection and as such are intrinsically subjective. However, it is clear that many of the new class I \choh maser detections are associated with GLIMPSE sources with ``green'' (i.e. 4.5\,$\mu$m excess) emission; see Figure \ref{fig:ch3oh_green} for an example source. \citet{cyganowski09} explains that the green emission band of GLIMPSE contains both H$_2$ and CO (1--0) lines, which are excited by shocks. Indeed, some of the new masers detected by MALT-45 have similar ``green'' emission as with EGOs, although less prominently so. This shows a close association between shocks traced by EGOs and class I \choh masers.

\subsection{95\,GHz class I \choh masers}
\label{sec:95_masers}
\citet{valtts00} found a relationship between the peak flux density of the 95 and 44\,GHz class I methanol masers (which come from the same transition family). They found that the peak flux density of 44\,GHz masers is approximately three times greater than the 95\,GHz masers.

\begin{figure}
  \includegraphics[width=0.47\textwidth]{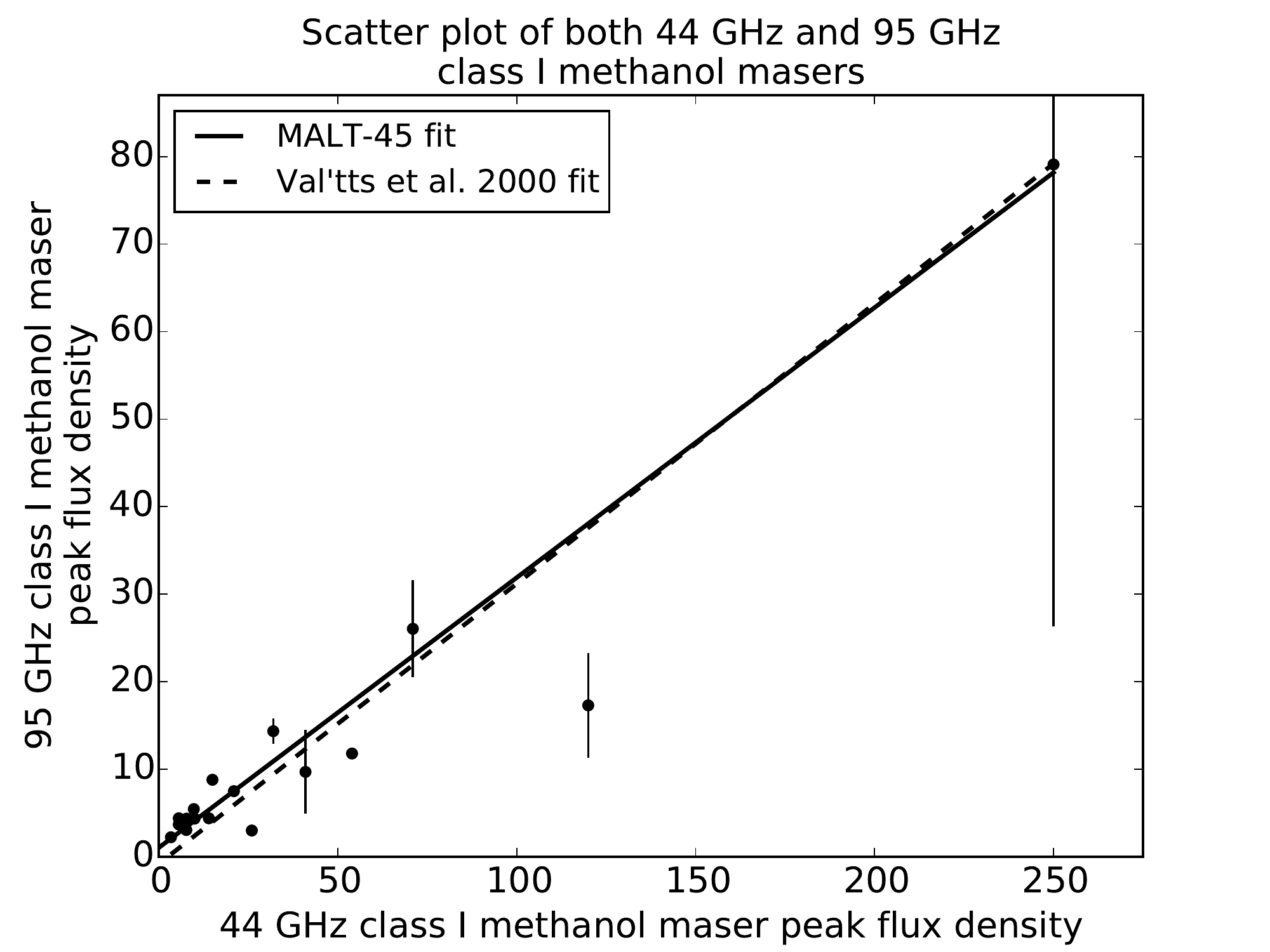}
  \caption{A scatter plot of 44\,GHz class I \choh masers from MALT-45 with known 95\,GHz class I \choh maser emission. There are 19 masers with both 44 and 95\,GHz emission. The line of best fit for these data is plotted alongside the line of best fit presented by \citet{valtts00}. Vertical lines indicate the minimum and maximum recorded values for 95\,GHz masers, if available, on top of the plotted average. Units of peak flux density are in Jy.}
  \label{fig:95ghz_comparison}
\end{figure}

Information for 95\,GHz class I \choh masers towards MALT-45 44\,GHz masers are obtained from \citet{valtts00,ellingsen05} and \citet{chen11}. In total, we have 19 class I maser sources with both 44 and 95\,GHz peak flux densities. When faced with more than one peak flux density over various epochs, we use the average of all values. Analysing the distribution in Figure \ref{fig:95ghz_comparison}, we find that our results agree with those of \citet{valtts00}. Our line of best fit is $P_{95}=0.31\times P_{44}+1.0$, with an $r^2$-value of 0.98, while the \citet{valtts00} fit is $P_{95} = 0.32\times P_{44}-8.1$, with an $r^2$-value of 0.53. The uncertainty of our fitted slope is $\pm0.01$. Note that our fit does not use the 44\,GHz maser with 120\,Jy; if we do include it, our fit becomes $P_{95}=0.28\times P_{44}+1.0$, with an $r^2$-value of 0.91 and uncertainty $\pm0.02$. This is not significantly different, but our slope closely agrees with \citet{valtts00} when the outlier is omitted.

\subsection{Class I \choh maser associations}
In this section, we discuss other maser species and transitions and their association with the class I \choh masers detected in MALT-45. Association with each maser is credited if the emission in question is within 60 arcsec of the class I maser position.

\subsubsection{Class II masers from the MMB}
\begin{figure}
  \includegraphics[width=0.47\textwidth]{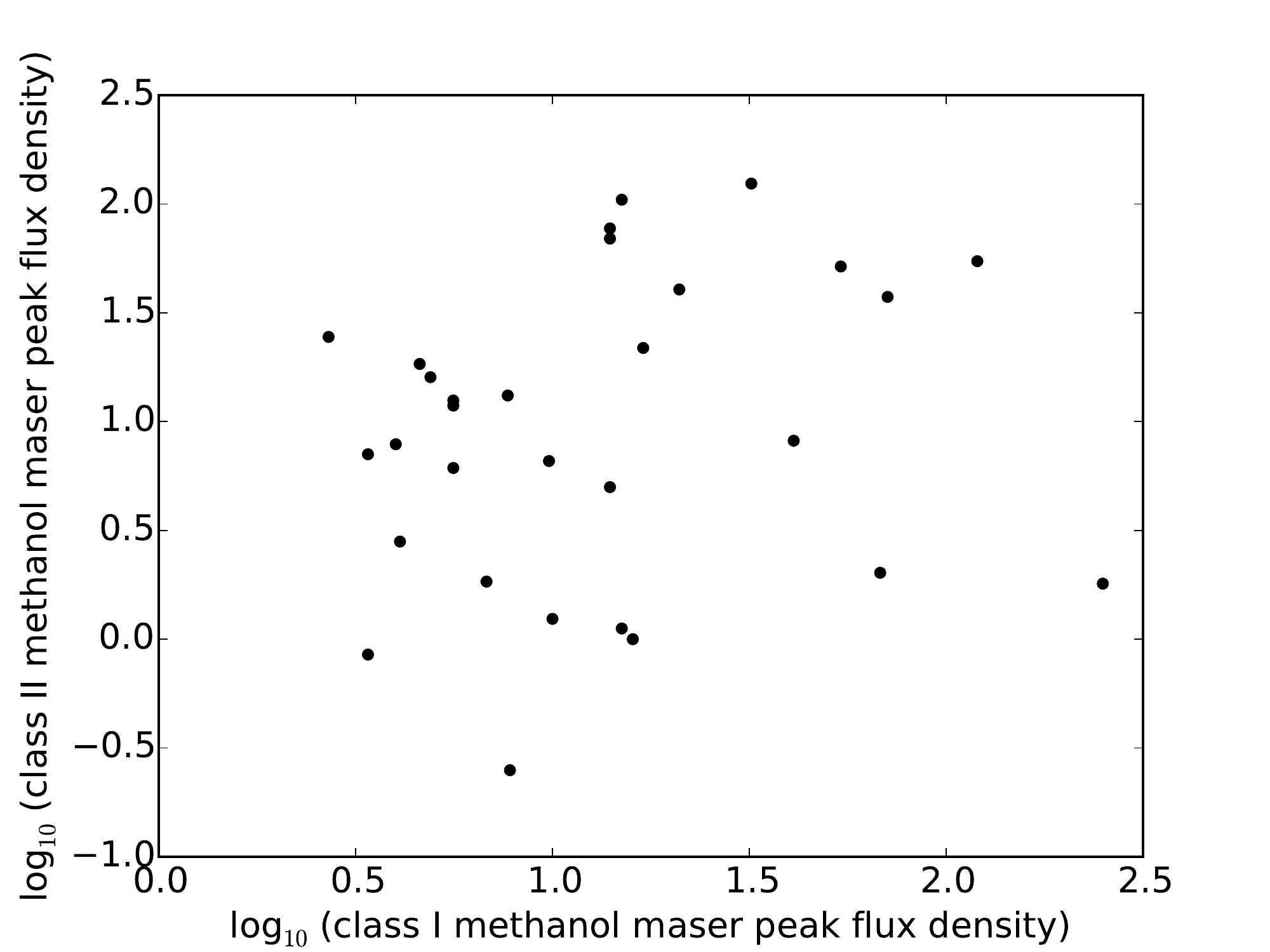}
  \caption{The luminosities of associated class I and class II methanol masers plotted against each other. No significant correlation is observed. Units of luminosity for the masers are Jy\,kpc$^{-2}$, using peak flux densities.}
  \label{fig:classII_maser_scatter_lum}
\end{figure}
%--------------------
\begin{figure}
  \includegraphics[width=0.47\textwidth]{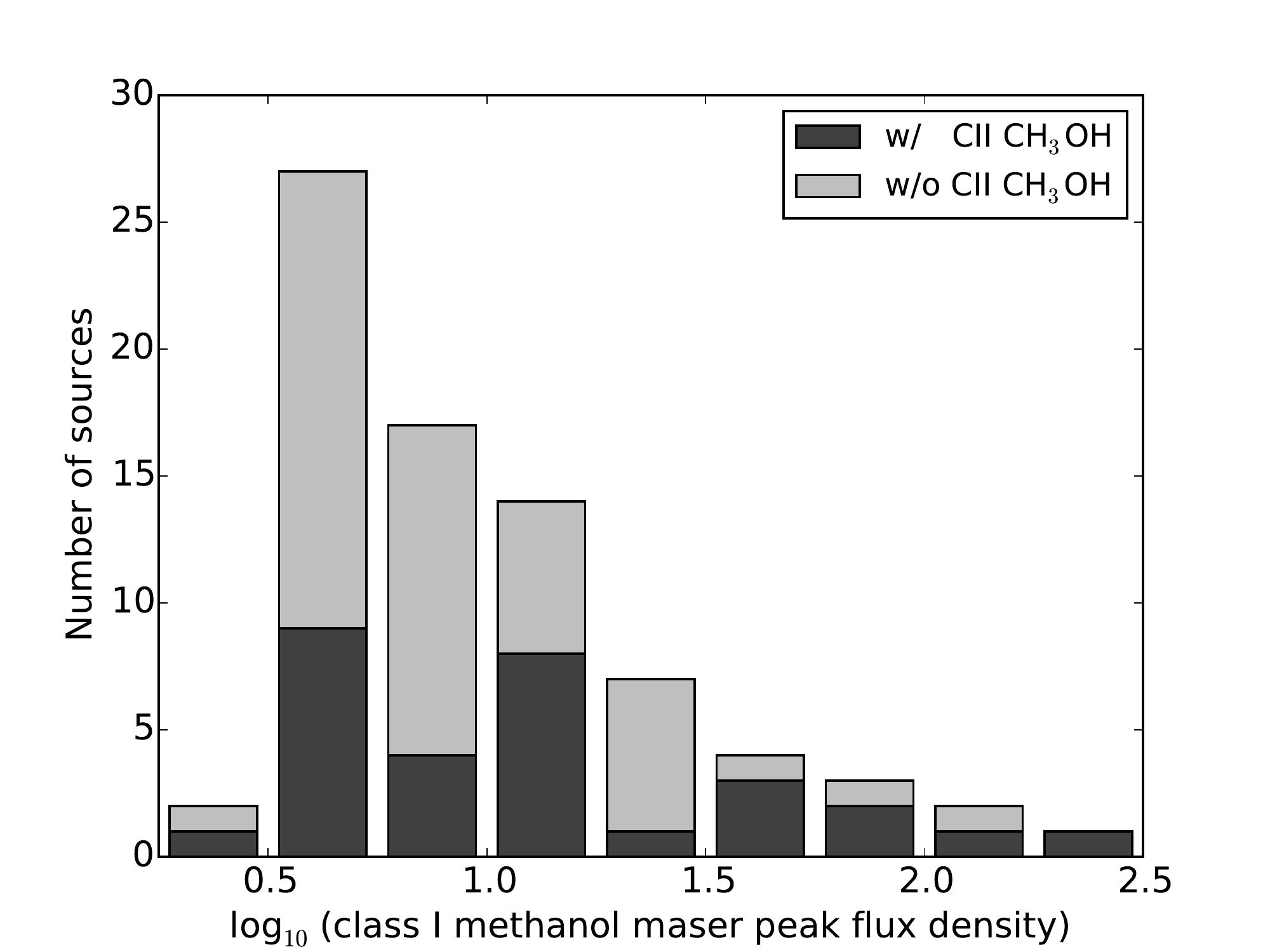}
  \caption{A stacked histogram of class I methanol masers with and without class II methanol masers. This plot shows that class II masers can be associated with a wide range of class I intensities.}
  \label{fig:classII_maser_hist}
\end{figure}

In the past, class I methanol masers have been identified from targeted searches towards class II maser sites \citep{ellingsen05}. For the first time, we are able to make an unbiased comparison of the rate at which each class of methanol maser occurs and their degree of overlap.

Within the MALT-45 survey region, 54 6.7\,GHz class II methanol masers have been detected by the MMB \citep{caswell11}. To compare against MALT-45 class I maser regions, 7 class II masers have been grouped together into ``clusters'', as they are close together (within 60 arcsec). The clusters containing grouped masers are G330.88$-$0.37, G331.54$-$0.07, G332.30$-$0.09, G333.13$-$0.44 and G333.23$-$0.06. While the velocities of each maser site grouped together may be different, there is significant overlap in the range of emission (not greater than 9\,\kms apart). Without high-resolution positions, inferring which class II maser is associated with the class I emission is difficult, and it is more practical to only compare their clustered position, especially as a class I \choh maser is almost certainly associated with the class II source. Therefore, we compare 47 class II methanol masers positions against the 77 class I sources detected in MALT-45. There are 28 class I \choh masers associated with the clusters, and 16 of these class I masers are new; see Table \ref{tab:ch3oh}. An association exception is made for G331.13$-$0.25, where the class II \choh maser is slightly further offset than 60 arcsec from the MALT-45 class I \choh maser position, but is well studied and known to be related \citep{ellingsen05,voronkov14}. Additionally, the next closest class I and class II \choh maser pair is approximately 100 arcsec, which aids to rule out mis-association. Thus, 60 per cent of class II masers have a class I counterpart (28/47), while approximately 36 per cent of class I masers have a class II counterpart (28/77).

\citet{ellingsen05} reports a 38 per cent association of class II sources with a 95\,GHz class I maser counterpart. Given that the 44\,GHz species is typically stronger than the 95\,GHz transition, and that the MALT-45 sensitivity is better, we perhaps expect our reported class II-to-class I association to be greater. As discussed in Section \ref{sec:95_masers}, \citet{valtts00} finds the relative peak flux density of 95\,GHz class I \choh masers to 44\,GHz class I \choh masers to be approximately one-third. Assuming \citet{ellingsen05} had a sensitivity cutoff for 95\,GHz masers at 3\,Jy, we can limit the results of the 44\,GHz population by a 9\,Jy cutoff. There are 17 44\,GHz masers with a peak flux density greater than 9\,Jy associated with a class II \choh maser (36 per cent), which is only slightly lower than the proportion of associations found by \citet{ellingsen05}. Therefore, our results are reasonably consistent. Considering that the MALT-45 sensitivity to class I \choh masers is better than that of \citet{ellingsen05} by approximately a factor of 2, and that we search for 44\,GHz masers instead of 95\,GHz masers, we are assured that there is a higher association rate than previously reported.

The luminosities of class I and class II \choh masers were compared, but no correlation was found ($r^2$-value of 0.03); see Figure \ref{fig:classII_maser_scatter_lum}. The relative populations for class I masers with and without a class II source are compared in Figure \ref{fig:classII_maser_hist}; the histogram reveals that class II sources are associated with a wide range of class I brightnesses, which reaffirms the lack of correlation. With sensitive, targeted 44\,GHz observations of class II masers, more data can be collected for a better analysis. Note that the search radius should be at least an arcmin to cover the range of separations between class I and class II masers; see Section 4.2 of \citet{voronkov14}.

\subsubsection{\water masers from HOPS}

\begin{figure}
  \includegraphics[width=0.47\textwidth]{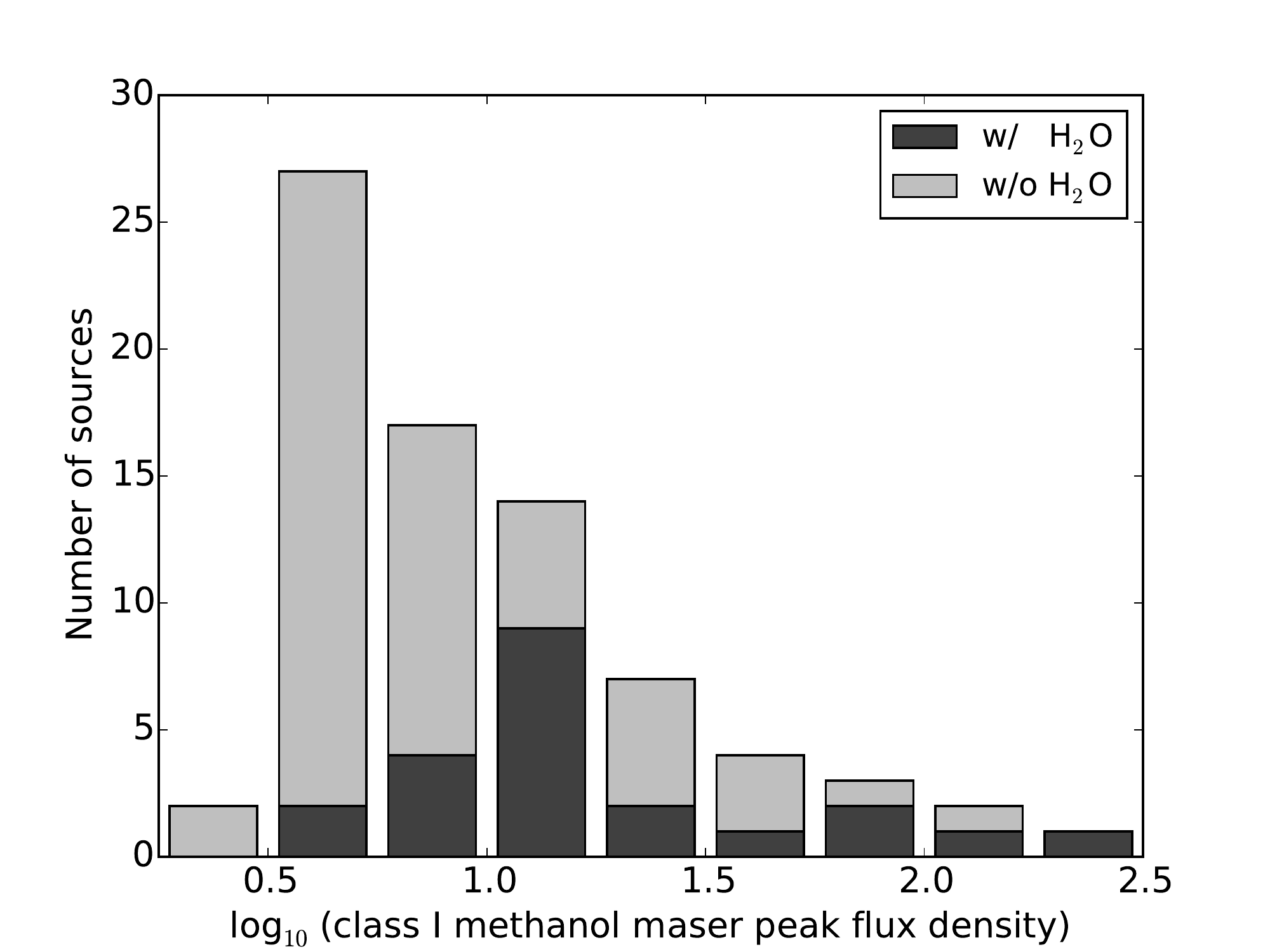}
  \caption{A stacked histogram of class I methanol masers with and without \water masers. This plot shows that \water masers tend to be associated with brighter class I methanol masers.}
  \label{fig:water_maser_hist}
\end{figure}

Water masers are another maser species that are collisionally excited. Therefore, we might expect a close association between class I methanol and water masers. HOPS has found \water masers in an untargeted way, and catalogues their findings into three association groups: star-forming, evolved star, and ``unknown'' \citep{walsh14}. \citet{breen10} also presents a catalogue of \water masers, typically associated with class II \choh masers and/or \oh masers.

HOPS has detected a total of 48 \water masers within the MALT-45 survey region \citep{walsh11,walsh14}. Of these, HOPS has classified 10 as being associated with an evolved star, leaving 38 as star-forming or ``unknown''. Including the \citet{breen10} detections brings the total to 71, but there is a large overlap with the HOPS catalogue, as well as many \water masers near to others (within 60 arcsec). Similar to the MMB class II \choh masers, after grouping masers into clusters, we compare MALT-45 class I \choh masers against a total of 43 \water maser clusters.

Of these 43 \water maser clusters, 27 are associated with a class I \choh maser (63 per cent). It is possible that the class I \choh population have more \water maser associations, but have not been found yet. \citet{breen10} targeted observations towards known sources, and HOPS has a lower-bound sensitivity between 5 and 10\,Jy. Follow-up observations for \water masers towards new class I \choh masers may be productive in identifying new \water masers.

In the HOPS catalogue recently published \citep{walsh14}, the water masers G331.86$+$0.06 and G333.46$-$0.16 were not redetected in the high-resolution follow-up survey. The G333.46$-$0.16 water maser is associated with a MALT-45 class I \choh maser, but this was a known class I \choh maser region (along with other maser species). The G331.86$+$0.06 water maser, however, is slightly offset from a new class I \choh maser. The offset is enough to not be included as an association, but it lends credence that this variable water maser is likely highlighting star-forming activity.

\subsubsection{OH masers from \citet{sevenster97} and \citet{caswell98}}

\begin{figure}
  \includegraphics[width=0.47\textwidth]{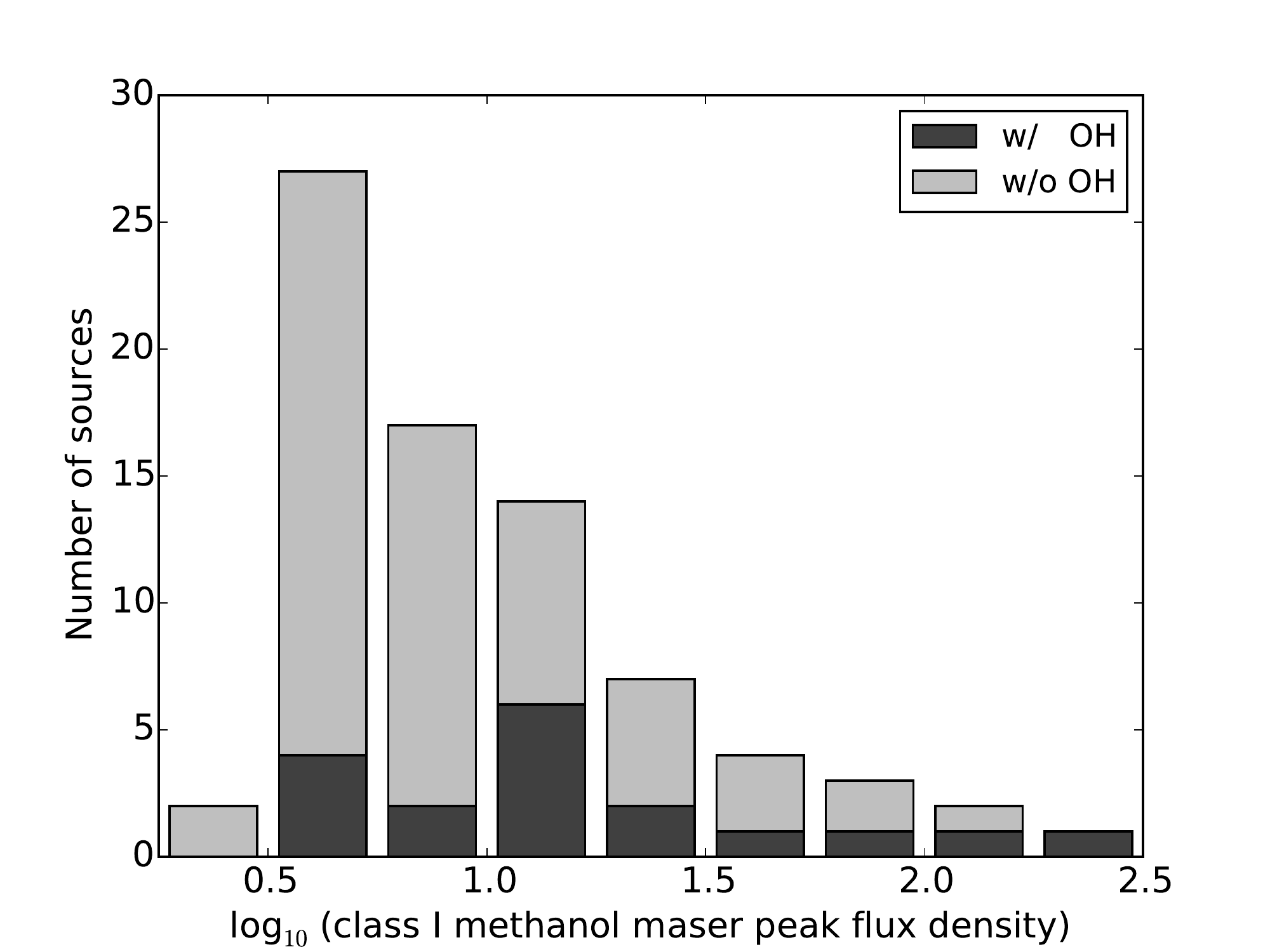}
  \caption{A stacked histogram of class I methanol masers with and without \oh (1612, 1665, or 1667\,MHz) masers. \oh masers appear to be associated with a wide range of class I methanol masers.}
  \label{fig:oh_maser_hist}
\end{figure}

\citet{voronkov10a} discuss the positioning of class I methanol maser emission on a HMSF evolutionary timeline, especially compared to other masers. Class II methanol masers have been characterised well in terms of their position in a HMSF timeline, associated with millimetre sources, but typically not \hii regions. Unlike class II masers, class I masers are currently thought to be associated with both early- and late-type HMSF. This stems from finding class I masers associated with outflows, but also with \oh masers.

To date, no unbiased comparison between \oh and class I methanol masers has been performed. With the first unbiased sample of class I masers, we are able to analyse the relative populations and perhaps refine class I masers on an evolutionary timeline. \citet{voronkov10a} mentions that class I masers associated with \oh masers are known primarily due to the class II methanol maser emission present, as most class I sources have been found toward class II sources. \citet{caswell97} discuss how \oh favoured maser regions have a greater number of associated \hii regions than class II masers, indicating that \oh masers typically signpost later stages of an evolutionary timeline than class II masers. Combining the \oh masers found by \citet{sevenster97} and \citet{caswell98}, there are 16 star-forming \oh masers within the MALT-45 survey region. The other \oh masers in \citet{sevenster97} and \citet{caswell98} are associated with evolved \oh/IR stars, determined from \textsc{SIMBAD} and individual source descriptions within \citet{caswell98}. 94 per cent (15/16) of star-forming \oh masers are associated with a class I \choh maser. Of these 15, only 2 masers are without a 6.7\,GHz class II methanol maser (13 per cent; G331.50$-$0.10 and G333.59$-$0.21). It is difficult to contrast maser regions containing \oh but not class II methanol, as almost every \oh maser in the survey region is associated with a class II \choh maser. If a population of \oh masers without class II \choh was isolated, a better comparison with class I \choh could be established. The MALT-45 survey results merely indicate that class I \choh is associated with all maser species found.

There are 30 class I \choh maser regions with associated class II \choh or \oh maser emission, indicating late stages of star formation. The remaining 47 masers are unaccounted for; if these are all associated with early-type star-formation, then class I \choh masers appear to be more closely associated with earlier stages in an evolutionary timeline.

\subsubsection{Summary of star-formation maser associations}
\begin{figure}
  \includegraphics[width=0.47\textwidth]{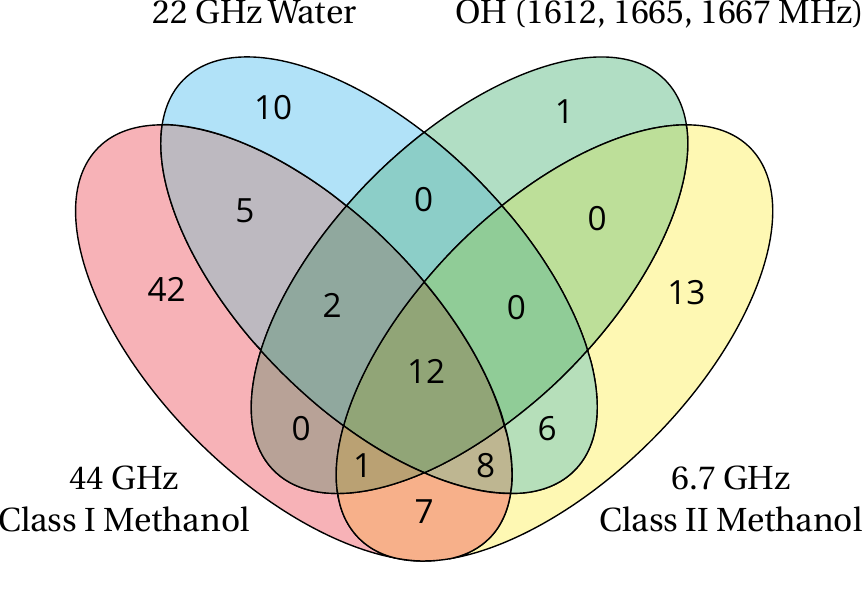}
  \caption{Venn diagram of all discussed maser region associations. The majority of maser regions have a class I \choh component, particularly with class II \choh as well, but there is a large population of class I masers without any associations.}
  \label{fig:maser_venn}
\end{figure}

A Venn diagram of the maser associations can be seen in Figure \ref{fig:maser_venn}. There appears to be a strong overlap between class I and class II \choh masers, with and without the presence of \water and/or \oh masers. This may suggest that the majority of class II \choh masers are associated with class I \choh masers because of an abundance of \choh. The smaller population of class II \choh masers without class I \choh emission (40 per cent; 19/47) may indeed have no association, or even weaker class I \choh emission than is detected by MALT-45.

There is a large population of class I \choh masers without any other maser counterpart. As discussed, these may highlight non-star-forming phenomena, young regions of HMSF, or low mass star-forming regions. However, given the strong association with \cs seen in Figure \ref{fig:cs_thrumms} and with other HMSF masers seen in Figure \ref{fig:maser_venn}, the majority of class I \choh masers are likely associated with HMSF.

Figures \ref{fig:classII_maser_hist}, \ref{fig:water_maser_hist} and \ref{fig:oh_maser_hist} show the luminosity population of class I \choh masers as stacked histograms. Generally, the presence or absence of other maser transitions does not appear to correlate with the class I \choh maser luminosity.

\citet{titmarsh14} discuss the placement of \water masers within an evolutionary timeline for HMSF, and concluded that they are not likely to be associated with any specific phase of a timeline, but rather dependant on the environment. As \water masers are collisionally excited like class I \choh masers, and class I masers are associated with a wide range of other species (with and without class II \choh, \oh), we tend to agree that the environment plays perhaps the predominant role in determining the likelihood of a region having an associated class I \choh maser rather than the evolutionary phase.

Investigation of the peak flux densities and luminosities of class I \choh masers versus class II \choh, \water and \oh masers does not reveal any relation. Investigation will be performed again in a subsequent MALT-45 maser follow-up paper, where maser regions can be better associated and flux density calibration is able to be more accurately determined.

\subsection{SiO $v=1,2,3$ maser associations}
\label{sec:sio_discussion}
MALT-45 has detected \siototal regions with at least one of the vibrationally excited lines $v=1,2,3$ of \sio (1--0) maser emission. Of these, 4 are associated with a \oh maser (all 1612\,MHz, G331.60$-$0.14 has an additional 1665\,MHz), and 8 are associated with a \water maser. \oh and \water masers are common companions with evolved infrared stars \citep{caswell98,walsh14}, and indeed GLIMPSE reveals that all \sio maser regions have an infrared star associated within 30 arcsec of the peak maser position. There is no overlap of \sio masers associated with both an \oh and \water maser. The population presented in this paper associated with \oh and \water masers is insufficient to perform correlation tests upon. None of the \sio masers detected have been reported previously; thus, MALT-45 is a productive survey for identifying evolved stars not found in other searches.

The \sio $v=1$ line generally appears to be the strongest of the three maser lines, and is the most common detection. There are eleven cases where the $v=2$ line is stronger than $v=1$, and two regions contain only $v=2$ emission. This result is somewhat unexpected, as generally the lower $v$ transitions should be brighter than the higher $v$ transitions. \citet{gray09} predict that \sio masers should have brightness decreasing with $v=1,2,3$, and the $v=2$ is often weaker due to overlap with a water line. Neither of the two regions with only \sio $v=2$ emission have other masers associated, but the sub-population of \sio regions with brighter $v=2$ than $v=1$ have three \water masers and two \oh masers associated. This is almost half of all masers associated with \sio regions. However, we currently cannot explain why some $v=2$ masers are stronger than their $v=1$ counterparts. Only three $v=3$ lines are detected, and these are all quite weak, especially compared to the $v=1,2$ lines within the same region; it appears that this line is difficult to detect without sensitive observations.

The only \sio maser with nearby \choh masers is G333.90$-$0.09, but is separated by 33 arcsec. The same \sio maser is 13 arcsec from a Sevenster 1612\,MHz OH maser, which is cospatial with a star in GLIMPSE. As the nearby class I and class II \choh masers are only 4 arcsec apart but separated from the star, we consider both pairs of masers to not be associated.

%%%%%%%%%%%%%%%%%%%%%%%%%%%%%%%%%%%%%%%%%%%%%%%%%%%%%%%%%%%%%%%%%%%%%%%%%%%%%%%%
\section{Summary and Conclusions}
MALT-45 has successfully probed the Galactic plane for \cs (1--0), 44\,GHz class I methanol masers, and the \sio (1--0) family of $v=0,1,2,3$ lines. This is the first survey to produce a large scale unbiased map of \cs, and is the first untargeted search of class I \choh masers and \sio emission. Further publications will reveal the full 7\,mm star-forming environment, through the remaining lines mapped by this survey. With time, MALT-45 could be extended to map more of the Galactic plane, revealing more about the nature of star formation within our Galaxy. With the current survey, we have determined:

(i) Across the survey region, the \cs/\nh ratio is low in the centres of many clumps, but higher at the edges. This is likely due to a combination of increased \nh abundance and \cs being depleted and/or becoming optically thick in the centres of clumps;

(ii) A large population of class I methanol masers are not associated with other masers (55\%);

(iii) Class I methanol masers are good indicators of systemic velocities;

(iv) Class I methanol masers are found towards most known class II methanol, water and hydroxyl masers within the survey region. Thus, positioning class I methanol on a maser timeline of HMSF is difficult, and the occurrence of class I methanol masers is perhaps more likely due to environmental factors;

(v) Silicon monoxide masers associated with evolved stars are most commonly detected in their $v=1$ vibrational mode. When a region contains more than one vibrational mode, the intensity typically descends with vibrational mode ($v=1,2,3$). However, we find two cases where we only detect $v=2$ and eleven cases where both $v=1$ and $v=2$ are detected, but the $v=2$ is stronger.

%%%%%%%%%%%%%%%%%%%%%%%%%%%%%%%%%%%%%%%%%%%%%%%%%%%%%%%%%%%%%%%%%%%%%%%%%%%%%%%%
\section*{Acknowledgements}
The authors sincerely thank the referee, Anita Richards, for timely and thorough comments that have improved the manuscript. C. H. Jordan thanks the staff of the Paul Wild Observatory for the many hours of assistance lent to the MALT-45 project. The Australia Telescope Compact Array is part of the Australia Telescope National Facility which is funded by the Commonwealth of Australia for operation as a National Facility managed by CSIRO. Shari Breen is the recipient of an Australian Research Council DECRA Fellowship (project number DE130101270). PJB acknowledges support from NSF grant AST-1312597 and the University of Florida. This research has made use of the SIMBAD database, operated at CDS, Strasbourg, France and NASA's Astrophysics Data System. \textsc{MIRIAD}\footnote{http://www.atnf.csiro.au/computing/software/miriad/}, \textsc{LiveData}, \textsc{Gridzilla}\footnote{http://www.atnf.csiro.au/computing/software/livedata/index.html} and \textsc{ASAP}\footnote{http://svn.atnf.csiro.au/trac/asap} are software packages managed and maintained by CSIRO Astronomy and Space Science.

%%%%%%%%%%%%%%%%%%%%%%%%%%%%%%%%%%%%%%%%%%%%%%%%%%%%%%%%%%%%%%%%%%%%%%%%%%%%%%%%

\end{document}